\documentclass[a4paper,11pt]{article}
\usepackage[utf8]{inputenc}
\usepackage{graphicx,xcolor}
\usepackage{setspace}
\usepackage{amssymb}
\usepackage{amsmath}
\newcommand{\dro}{\bar{r}_{\rm o}}

\newcommand{\dr}{\bar{r}}
\newcommand{\du}{\bar{u}}
\newcommand{\dPfi}{\bar{P}_{\rm fi}}
\newcommand{\dPf}{\bar{P}_{\rm f}}

\newcommand{\beq}{\begin{equation}}
\newcommand{\eeq}{\end{equation}}
\newcommand{\dd}{{\rm d}}

\usepackage[round]{natbib}
\RequirePackage{lineno}

\date{}

\title{Initiation of fluid-induced fracture in a thick-walled hollow permeable sphere}

\author{Peter Grassl$^{1*}$, Milan Jir\'{a}sek$^{2}$ and Domenico Gallipoli$^{3}$}

\begin{document}
\maketitle

$^{1}$ School of Engineering, University of Glasgow, peter.grassl@glasgow.ac.uk\\
$^{2}$ Department of Mechanics, Faculty of Civil Engineering, Czech Technical University in Prague, milan.jirasek@fsv.cvut.cz\\
$^{3}$ Laboratoire SIAME, Federation IPRA, University of Pau and Pays Adour, domenico.gallipoli@univ-pau.fr\\
$^{*}$ Corresponding author, Email: peter.grassl@glasgow.ac.uk

\section*{Abstract}
The initiation of fluid-induced fracture in formations of permeable geomaterials subjected to quasi-stationary flow processes (drained response) can be strongly affected by Biot's coefficient and the size of the formation.
The aim of this paper is to analyse the influence of these parameters on the initial fracture process of a thick-walled hollow permeable sphere subjected to fluid injection in the hole.
Assuming that fracture patterns are distributed uniformly during the hardening stage of the fracture initiation process, the coupled fluid-solid problem is described by a nonlinear ordinary differential equation, which is solved numerically by means of finite differences combined with shooting and Newton methods. The finite difference code has also been validated in the elastic range, i.e., before initiation of fracture, against an original closed-form analytical solution of the above differential equation.
The results show that the nominal strength of the sphere increases with increasing Biot's coefficient and decreases with increasing size.

Keywords: Fracture, hollow sphere, permeability, damage

\section{Introduction}
Interactions between fluid flow and fracture are important for processes resulting in the failure of flood defence embankments and earth or concrete dams \citep{SloSao00} but also in the deterioration of building materials, such as corrosion-induced cracking of reinforced concrete \citep{AndAloMol93} where the expansion of corrosion products in fluid form causes fracture in the material. These interactions are also important for the study of fluid-induced fracture processes in geological formations in the form of injection of sills \citep{Gou05} and clastic dykes \citep{Mee09}.
Recent research activities in fluid-induced fracture are driven by technologies such as hydraulic fracturing for unconventional oil and gas extraction \citep{GalReeHol07,GalLauOls14}, enhanced geothermal energy systems \citep{CheNarYan00} and underground storage of gas.

Examples of mathematical approaches to modelling the propagation of macroscopic cracks due to fluid injection include analytical models \citep{SavDet02,Det04,Det16}, finite element based solutions \citep{AdaSiePei07,CarGra12,MieMauTei15,LecDes15,WilLan16,deBVecOrt17,CaoHusSch18,VieGar18} and discrete approaches \citep{DamDetCun15,GraFahGal15}. Recent examples of experimental work are found in \cite{XinYosAda17}.
Initiation of hydraulic fracture close to a well-bore and the resulting tortuosity were investigated in \cite{AtkThi93,ZhaJefBun11}. Damage evolution close to boreholes in the form of borehole breakdown were studied experimentally in \cite{CusRutHol03,DreStaRyb10}. Damage and fracture initiation due to expansive pressures was treated in \cite{Lad67,Lec12,TarBlaFak16,GraFahGal15}. Experimental aspects of fluid-induced fracturing were studied in \cite{StaMaySha11}. Interactions between fluid flow and fracture play also an important role in many technologies outside the area of geomaterials \citep{KliRosKam16}.

In situations of material deterioration in which fluid pressure builds up internally over a very long period of time, the process of fluid-induced fracture can be modelled assuming quasi-stationary flow processes (drained response).
This was done in \cite{GraFahGal15}, where the effect of fluid pressure on elastic deformations and fracture initiation in a thick-walled cylinder was studied by means of a numerical network model. In this work, the elastic response from the network approach was compared with a closed-form analytical solution proposed in \cite{GraFahGal15}. In \cite{FahWheGal17}, the above analytical solution was extended and solved numerically to consider initiation of fracture during corrosion-induced cracking of reinforced concrete for the special case of zero Poisson's ratio and zero Biot's coefficient. These nonlinear analyses with zero Biot's coefficient are also similar to the mechanical approaches presented in \cite{YuHou91,PanPap01}. For most geomaterials, however, Biot's coefficient is not zero and is expected to have a significant effect on fracture initiation.

In the present study, we therefore extend the above analytical approaches to nonzero values of Biot's coefficient and Poisson's ratio. In particular, we present a poro-mechanics analysis of the fracture of a hollow thick-walled sphere subjected to inner fluid pressure for the full range of Poisson's ratios and Biot's coefficients assuming quasi-static flow processes (drained response). This work can be seen as an extension of the elastic solution of a material proposed by Lam\'{e} (e.g. \cite{TimGoo87}) by modelling the fracture process and considering the effect of fluid pressure on the solid \citep{Cou11}.
The adopted geometry of a hollow sphere is motivated by its frequent adoption in mathematical models for a wide range of processes. The case of spherical cavities in porous materials subjected to inner fluid pressure has been studied for biological processes of fluid injection \citep{BarAld92,AhmSidMah17}, the response of magma chambers in volcanology \citep{McT87}, ice formation in geology \citep{VlaWor10} and radioactive waste storage in civil engineering \citep{SelSuv14}.
  In many of these physical processes, fracture and damage play an important role, but were not included in the mathematical modelling.
  The new contribution of the present study is that a mathematical model for fluid-induced fracturing of a spherical permeable hollow sphere subjected to inner fluid pressure is proposed, which considers the influence of Biot's coefficient.

The presented approach is based on a number of simplifications. Spherical symmetry is assumed for the elastic response. For the fracture response, a regular arrangement of fracture patterns is assumed for the initial (hardening) response. In the post-peak regime, cracks are usually localised, so that the assumption of a regular arrangement of fracture patterns is not valid anymore. The effect of fracture on transport properties is assumed to be small so that that the permeability and Biot's coefficient are taken to be constant across the sphere and throughout the loading process. Furthermore, the fluid is considered as incompressible and of constant viscosity. Variations of the rate at which the fluid is injected into the hole of the sphere are so slow that stationary flow conditions prevail and a drained response is obtained. For elasticity, more complicated cases considering fast rates are discussed in \citet{Che16}.
Finally, displacements are assumed to be small and not influenced by gravity.

The paper is divided into four parts. Firstly, the fluid-driven loading is defined and the pressure distribution across the sphere is calculated in Section~\ref{sec:flow}. 
Then, the model of the elastic response of the sphere is described and a closed-form analytical solution is derived in Section~\ref{sec:elastic}. The elastic response is then extended to nonlinear fracture mechanics in Section~\ref{sec:fracture}, where the effect of Biot's coefficient and size on the nominal strength of the sphere is also studied. 

\section{Fluid-driven loading and pore pressure distribution}\label{sec:flow}

\begin{figure}
  \begin{center}
    \begin{tabular}{cc}
      \includegraphics[width=7cm]{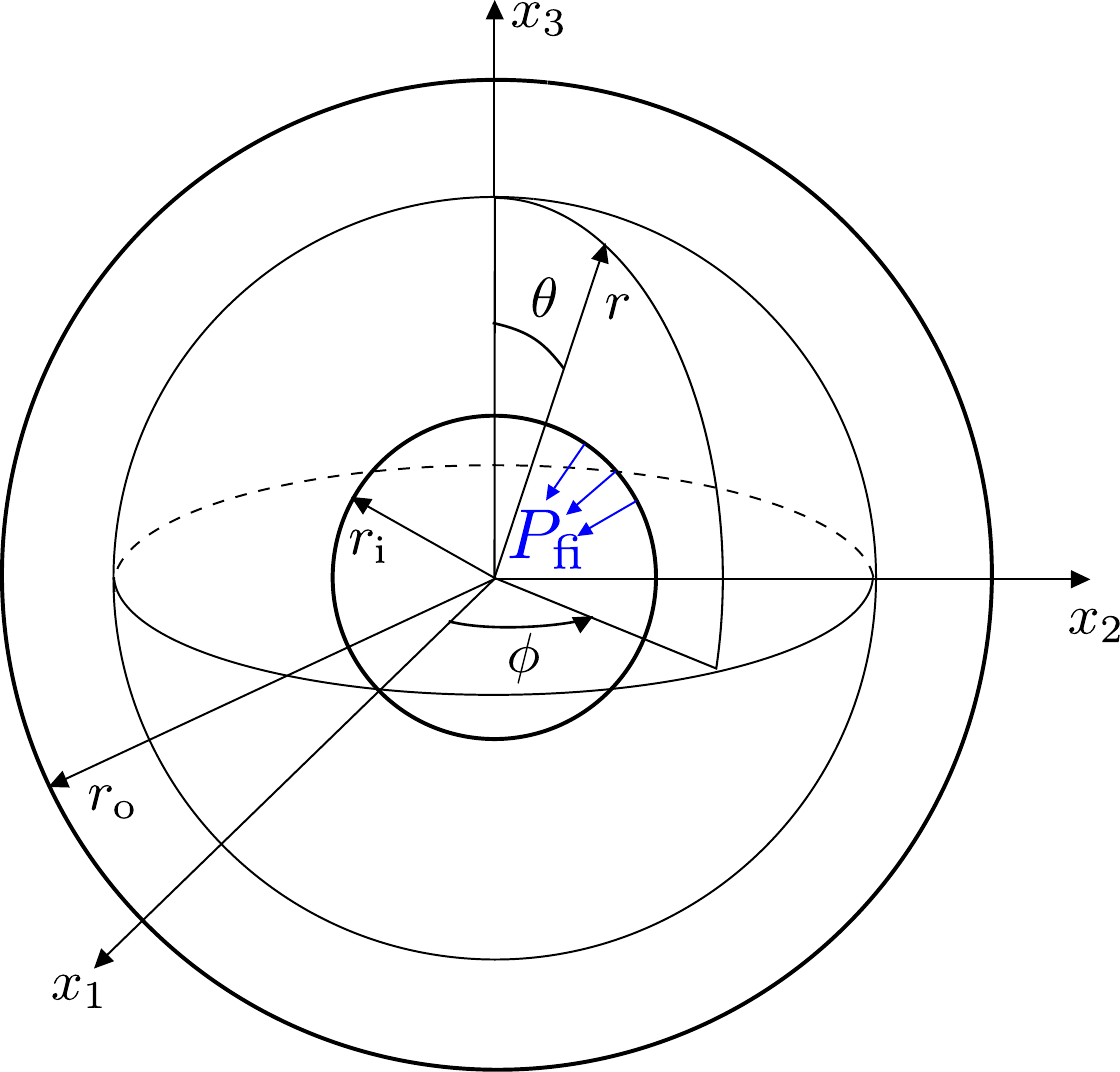} & \includegraphics[width=6cm]{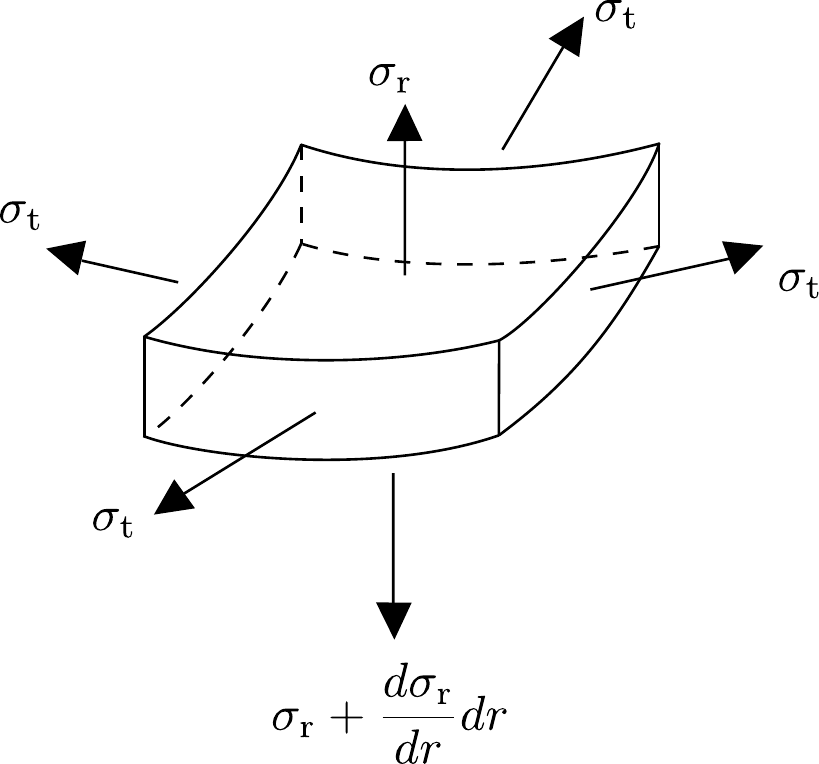}\\
      (a) & (b)
    \end{tabular}
  \end{center}
  \caption{Thick-walled hollow sphere: (a) geometry and coordinate system and (b) stresses acting on a small element in the sphere \cite{TimGoo87}.}
  \label{fig:thickSphere}
\end{figure}

In the present section, the analytical solution of the fluid pressure distribution and the mechanical response of a thick-walled hollow sphere subjected to internal fluid pressure under steady-state conditions is presented (Figure~\ref{fig:thickSphere}a).
The hydraulic loading process is modelled as an increase of the incompressible fluid volume in the hole inside the sphere. Part of this increase of volume is accommodated by an expansion of the inner hole of the sphere, and the remaining part of the fluid volume flows through the permeable sphere.
The volume balance is described by
\begin{equation} \label{eq:flow1}
  \dot{V} = \dot{V}_{\rm i} +  Q
\end{equation}
where $Q$ is the total fluid volume flow through the inner boundary of the hollow sphere and 
\begin{equation} \label{eq:flow2}
\dot{V}_{\rm i}  =  4 \pi r_{\rm i}^2\dot{u}_{\rm i}
\end{equation} 
is the rate of volume increase of the hole, expressed as the product
of the inner surface area, $4 \pi r_{\rm i}^2$, and the 
displacement rate at the inner surface, $\dot{u}_{\rm i}$.
For small displacements, the inner radius $r_{\rm i}$ can be considered
as constant (for the purpose of surface area evaluation).

The fluid in the hole is under pressure $P_{\rm fi}$ and the pressure gradient induces flow of the fluid through the permeable sphere.
It is assumed that the fluid is incompressible with constant viscosity.
Furthermore, the sphere is fully saturated and possesses a constant permeability.
The flow is also considered to satisfy steady-state conditions, which makes the fluid flux time-independent.

From the total flow rate $Q$, the fluid pressure distribution across the sphere can be determined.
Imposing conservation of fluid mass, combined with the assumption of radial symmetry and fluid incompressibility, one can infer that the tangential flow vanishes and that the total flow rate through any concentric spherical surface is the same, independent of the surface radius, $r$.
Consequently, the radial flux $q$ (radial volume flow rate per unit area) at a given distance $r$ from the centre of the thick-walled hollow sphere is calculated as 
\begin{equation} \label{eq:flow3}
q(r) = \dfrac{Q}{4\pi r^2}
\end{equation}

The radial flux is assumed to be linked to the
fluid pressure gradient by Darcy's law 
\begin{equation}\label{eq:flow4}
q(r) = \dfrac{\kappa}{\mu} \dfrac{{\rm d}P_{\rm f}(r)}{{\rm d} r}
\end{equation}  
where $\kappa$ is the intrinsic permeability [m$^2$] and $\mu$ is the dynamic shear viscosity of the fluid [Pa$\cdot$s].
The sign convention adopted here is that positive pore fluid pressure 
$P_{\rm f}$ corresponds to tension (i.e., the actual values of $P_{\rm f}$ are negative).

By setting the right-hand sides of (\ref{eq:flow3}) and (\ref{eq:flow4}) equal and then integrating, we obtain
\begin{equation}\label{eq:flow5}
P_{\rm f}(r) = - \dfrac{\mu Q}{4 \kappa \pi r}  + C
\end{equation}
Here, $C$ is an integration constant, which is determined from a boundary
condition. It is assumed that fluid pressure at the outer boundary 
(spherical surface of radius $r_{\rm o}$) vanishes, i.e., 
$P_{\rm f}(r_{\rm o}) = 0$, which leads to
\begin{equation}\label{eq:flow6}
C = \dfrac{\mu Q}{4 \kappa \pi r_{\rm o}}
\end{equation}
Recall that the fluid pressure at the inner boundary  
(spherical surface of radius $r_{\rm i}$) has already been denoted as
$P_{\rm fi}$. By imposing $P_{\rm f}(r_{\rm i}) = P_{\rm fi}$,
we can express the total flux 
\begin{equation} \label{eq:flow8}
Q =  P_{\rm fi} \dfrac{4 \kappa \pi r_{\rm i} r_{\rm o}}{\mu \left(r_{\rm i} - r_{\rm o}\right)}\end{equation}
in terms of the inner pressure and construct the final formula for pore pressure distribution,
\begin{equation}\label{eq:flow9}
P_{\rm f}(r) =  P_{\rm fi} \dfrac{r_{\rm i}}{r_{\rm i} -r_{\rm o}}\dfrac{r - r_{\rm o}}{r}
=  P_{\rm fi} \dfrac{r_{\rm i}/r_{\rm o}}{r_{\rm i}/r_{\rm o} -1}\dfrac{r/r_{\rm o} - 1}{r/r_{\rm o}}
\end{equation}
Note that the pore pressure depends on $r$, $r_{\rm i}$, $r_{\rm o}$ and $P_{\rm fi}$, but is independent of the 
intrinsic permeability $\kappa$ and absolute (dynamic) viscosity of the fluid $\mu$, as long as they are constant across the thickness of the sphere.

\begin{figure}
  \centering
  \includegraphics[width=12cm]{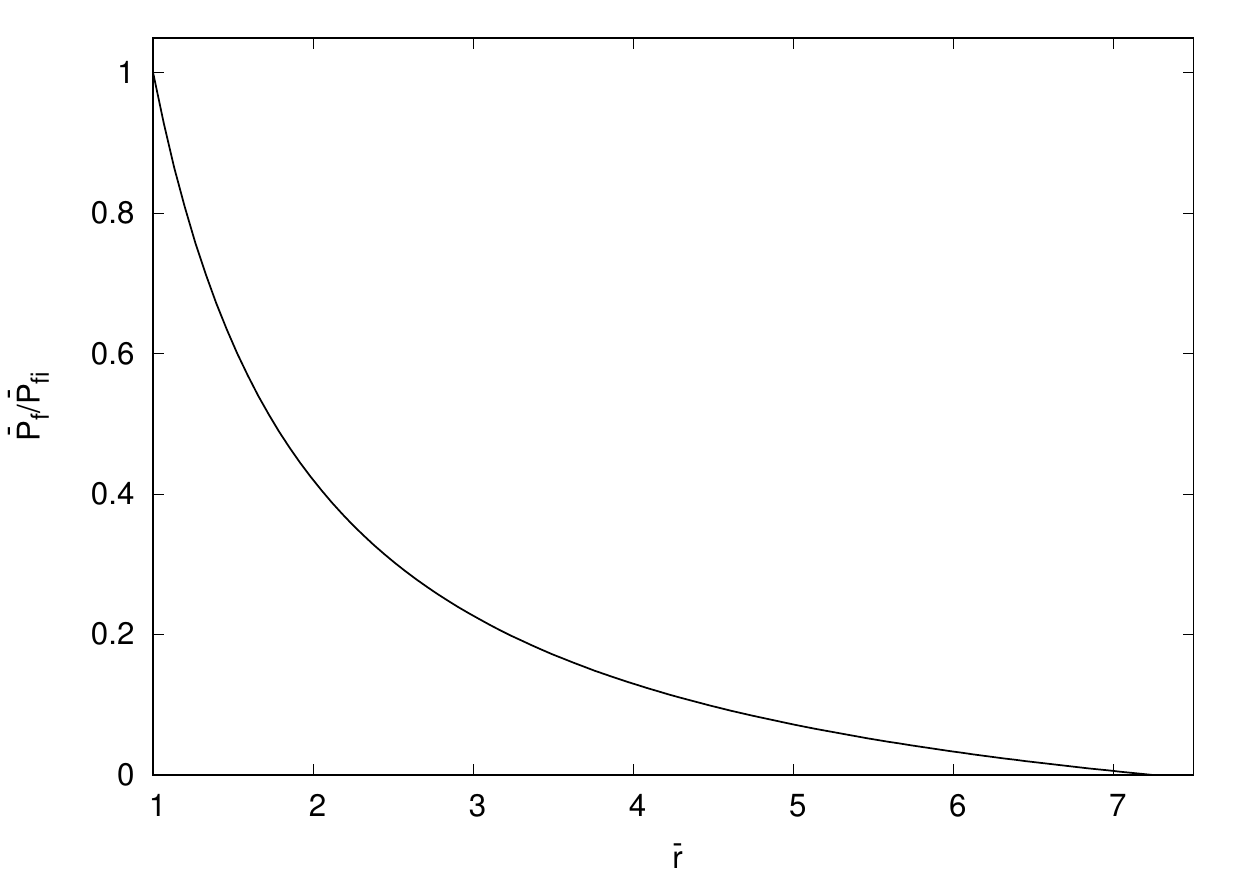}
  \caption{Distribution of normalised {\bf fluid pressure} plotted as function of dimensionless radial coordinate for $\dr_{\rm o}=7.25$.}
  \label{fig:flow}
\end{figure}

It is convenient to introduce dimensionless variables $\bar{r} = r/r_{\rm i}$, $\bar{r}_{\rm o} = r_{\rm o}/r_{\rm i}$, $\dPf = P_{\rm f}/E$ and $\dPfi = P_{\rm fi}/E$, where $E$ is Young's modulus of the porous material. 
In dimensionless form,
(\ref{eq:flow9}) is rewritten as
\begin{equation} \label{eq:dimFlow}
\dPf(\bar r) = \dPfi \dfrac{\dro - \dr}{\left(\dro-1\right)\dr}
\end{equation}
This is graphically illustrated in Figure~\ref{fig:flow}, which shows the normalised pore pressure $\dPf/\dPfi$ as function of the dimensionless radial coordinate $\dr$ (plotted for $\dr_{\rm o}=7.25$).

\section{Linear Elastic Response} \label{sec:elastic}

The mechanical response of the thick-walled hollow sphere due to fluid injection described in Section~\ref{sec:flow} is initially investigated for a linear elastic material.
In section~\ref{sec:elasticDerivation}, the equations for the linear elastic response are derived. Then, in section~\ref{sec:elasticResults} the results for varying Biot's coefficient and Poisson's ratio are presented.

\subsection{Derivation of equations for linear elastic response}\label{sec:elasticDerivation}
In this section the equations for the elastic response are derived.
The equilibrium equation of the hollow thick-walled sphere under spherical symmetry conditions (Figure~\ref{fig:thickSphere}b) was derived e.g.\ in \citet{TimGoo87} in the form
\begin{equation}\label{eq:elastic1}
\dfrac{\dd\sigma_{\rm r}}{\dd r} + 2\dfrac{\sigma_{\rm r} - \sigma_{\rm t}}{r} = 0
\end{equation}
where $\sigma_{\rm r}$ and $\sigma_{\rm t}$ are the total radial and tangential stresses, respectively, which are also the principal stresses, due to radial symmetry. Note that the tangential stress $\sigma_{\rm t}$ corresponds to two identical circumferential stresses as shown in Figure~\ref{fig:thickSphere} (i.e. $\sigma_{\rm t}$ = $\sigma_{\rm \phi} = \sigma_{\rm \theta}$). 
In poroelasticity, the total radial and tangential stresses $\sigma_{\rm r}$ and $\sigma_{\rm t}$ are equal to the sum of effective (mechanical) stresses, $\sigma_{\rm r}^{\rm m}$ and $\sigma_{\rm t}^{\rm m}$, and a certain multiple of the pore fluid pressure, $P_{\rm f}$. In the present notation (tension positive for stresses as well as pressure), we write $\sigma_{\rm r} = \sigma_{\rm r}^{\rm m} + b P_{\rm f}$ and $\sigma_{\rm t} = \sigma_{\rm t}^{\rm m} + b P_{\rm f}$ where $b$ is Biot's coefficient ranging between 0 and 1.
In this work, Biot's coefficient is interpreted as $b=1-K_d/K_s$ where $K_s$ is the macroscopic bulk modulus of the material at drained conditions and $K_s$ is the bulk modulus of the material that forms the solid skeleton between fluid accessible pores \citep{DetChe95,Cou11}. For $b \rightarrow 0$, one gets $K_d \rightarrow K_s$, which is only possible if the fluid accessible porosity tends to zero.

Substituting the expression of the total stresses into (\ref{eq:elastic1}), the equilibrium equation expressed in terms of effective stresses and fluid pressure is obtained:
\begin{equation} \label{eq:elastic2}
\dfrac{\dd\sigma_{\rm r}^{\rm m}}{\dd r} + 2 \dfrac{\sigma_{\rm r}^{\rm m} - \sigma_{\rm t}^{\rm m}}{r} +  b \dfrac{\dd P_{\rm f}}{\dd r} = 0 
\end{equation}
Combining this equilibrium equation with the strain-displacement equations and the elastic constitutive law, we will construct a differential equation from which the displacement field can be evaluated.

Under radial symmetry, the radial and tangential strains, $\varepsilon_{\rm r}$ and $\varepsilon_{\rm t}$, are linked to the radial displacement $u$ by the kinematic equations
\begin{eqnarray} \label{eq:elastic3}
  \varepsilon_{\rm r} &=& \dfrac{\dd u}{\dd r}
\\
\label{eq:elastic4}
  \varepsilon_{\rm t} &=& \dfrac{u}{r}
\end{eqnarray} 
If the material is linear elastic and isotropic, the constitutive equations (for the given triaxial stress state with two equal principal stresses) read
\begin{eqnarray} \label{eq:elastic5x}
\sigma_{\rm r}^{\rm m} &=& \dfrac{E}{(1-2\nu)(1+\nu)}\left((1-\nu)\varepsilon_{\rm r}+2\nu\varepsilon_{\rm t}\right)
\\
\label{eq:elastic6x}
\sigma_{\rm t}^{\rm m} &=& \dfrac{E}{(1-2\nu)(1+\nu)}\left(\nu\varepsilon_{\rm r}+\varepsilon_{\rm t}\right)
\end{eqnarray} 
where  $E$ is Young's modulus and $\nu$ is Poisson's ratio of the permeable material.

Combining the kinematic equations (\ref{eq:elastic3})--(\ref{eq:elastic4}) with the elastic constitutive law (\ref{eq:elastic5x})--(\ref{eq:elastic6x}) and substituting into the equilibrium condition (\ref{eq:elastic2}),
we obtain a differential equation for the radial displacement $u$ in the form
\begin{equation} \label{eq:elastic7}
  \dfrac{\dd^2 u}{\dd r^2} + 2\dfrac{\dd u}{\dd r}\dfrac{1}{r}-2\dfrac{u}{r^2} + b  \dfrac{P_{\rm fi}}{E} \dfrac{\left(1+\nu\right)\left(1-2\nu\right)}{\left(1-\nu\right)}\dfrac{r_{\rm i} r_{\rm o}}{r_{\rm i}-r_{\rm o}} \dfrac{1}{r^2} = 0
\end{equation}
In terms of the dimensionless variables introduced in Section~\ref{sec:flow} and the additional dimensionless variable $\du = u/r_i$, equation (\ref{eq:elastic7}) reads
\begin{equation} \label{eq:elastic8}
\dfrac{\dd^2\du}{\dd\dr^2} + 2 \dfrac{\dd\du}{\dd\dr}\dfrac{1}{\dr}-2\dfrac{\du}{\dr^2} + b \dPfi \dfrac{\left(1+\nu\right)\left(1-2\nu\right)}{\left(1-\nu\right)} \dfrac{\dro}{1-\dro} \dfrac{1}{\dr^2} = 0
\end{equation}
This second-order differential equation differs from the standard one for linear elastic materials in \cite{TimGoo87} because of the term involving Biot's coefficient. For the linear elastic constitutive law, equation (\ref{eq:elastic8}) is solved here both analytically in closed-form and numerically by using a finite difference scheme. 
The main steps of the closed-form solution are outlined next while the details of the numerical solution are presented in Appendix~\ref{app:numerical}.

The general solution of the differential equation (\ref{eq:elastic8}) is given by
\begin{equation}\label{eq:elastic9}
\du(\dr) = \dfrac{1}{2} b \dPfi \dfrac{\left(1+\nu\right)\left(1-2\nu\right)}{\left(1-\nu\right)}\dfrac{\dro}{1-\dro} + \dfrac{C_1}{\dr^2} + C_2 \dr
\end{equation}
and contains two integration constants $C_1$ and $C_2$ that need to be determined from boundary conditions.
At the inner boundary, the total radial stress is imposed to reflect the application of the fluid pressure, i.e. $\sigma_{\rm r}(r_{\rm i})=P_{\rm fi}$.  At the outer boundary, various hypotheses can be made and, in the present work, we assume that no stress is applied, i.e. $\sigma_{\rm r}(r_{\rm o})=0$. 

We next recall that $\sigma_{\rm r}=\sigma_{\rm r}^{\rm m} + b P_{\rm f}$
and that the values of pore pressure at the inner and outer boundaries are respectively equal to $P_{\rm f}(r_{\rm i})=P_{\rm fi}$ and $P_{\rm f}(r_{\rm o})=0$. This means that the two boundary conditions can be rewritten in terms of effective stresses as $\sigma_{\rm r}^{\rm m}(r_{\rm i})=(1-b)P_{\rm fi}$ and $\sigma_{\rm r}^{\rm m}(r_{\rm o})=0$, which can be further expressed in terms of radial displacement and its derivative by making use of the constitutive law (\ref{eq:elastic5x}) and kinematic equations (\ref{eq:elastic3})--(\ref{eq:elastic4}). After conversion to dimensionless form, the boundary conditions at the inner and outer boundaries are expressed as
\begin{eqnarray}\label{eq:elastic100}
(1-\nu)\dfrac{\dd \du(1)}{\dd \dr}+2\nu\du(1) &=& (1-b) (1-2\nu)(1+\nu) \bar{P}_{\rm fi} 
 \\
 \label{eq:elastic101}
 (1-\nu)\dfrac{\dd \du(\dr_{\rm o})}{\dd \dr}+2\nu\dfrac{\du(\dr_{\rm o})}{\dr_{\rm o}} &=& 0
\end{eqnarray}
where the dimensionless inner radius $\dr_{\rm i}=1$.

Substituting the general solution (\ref{eq:elastic9}) into (\ref{eq:elastic100})--(\ref{eq:elastic101}), we obtain a set of two linear equations from which the two integration constants 
\begin{eqnarray} \label{eq:elastic114}
C_1 &=& - \dPfi \left(1-b \dfrac{1-2\nu}{1-\nu}\right) \dfrac{\dro^3}{\dro^3-1} \dfrac{1+\nu}{2}
\\
\label{eq:elastic115}
  C_2 &=& \dPfi \left(1-2\nu\right) \left[ \left(1-b\dfrac{1-2\nu}{1-\nu}\right) \dfrac{1}{1-\dro^3} - \dfrac{b\nu}{1-\nu} \dfrac{1}{1-\dro}\right]
\end{eqnarray}
are easily evaluated.
The particular solution satisfying the given boundary conditions 
 ($\sigma_{\rm r}(r_{\rm i})=P_{\rm fi}$ and $\sigma_{\rm r}(r_{\rm o})=0$)
is therefore given by
\begin{equation} \label{eq:elastic116}
\du(\dr) = -\dPfi \left[\left(1- b \dfrac{1-2\nu}{1-\nu}\right) \dfrac{1}{\dro^3 -1} \left(\dfrac{1+\nu}{2} \dfrac{\dro^3}{\dr^2} + \left(1-2\nu\right) \dr \right) + b \dfrac{1-2\nu}{1-\nu} \dfrac{1}{\dro-1} \left(\dfrac{1+\nu}{2} \dro - \nu \dr\right)\right]
\end{equation}
and the resulting dimensionless effective stresses are
\begin{eqnarray}  \label{eq:elastic119}
\bar{\sigma}_{\rm r}^{\rm m}(\dr) &=& \dfrac{\sigma_{\rm r}^{\rm m}(\dr)}{E}= \dPfi \left[\left(1-b\dfrac{1-2\nu}{1-\nu}\right) \dfrac{1}{\dro^3 -1} \left(\dfrac{\dro^3}{\dr^3}-1\right) - b \dfrac{\nu}{1-\nu}\dfrac{1}{\dro-1}\left(\dfrac{\dro}{\dr}-1\right)\right]
\\
\label{eq:elastic120}
 \bar{\sigma}_{\rm t}^{\rm m}(\dr) &=&\dfrac{\sigma_{\rm t}^{\rm m}(\dr)}{E}=  - \dPfi \left[\left(1-b\dfrac{1-2\nu}{1-\nu}\right) \dfrac{1}{\dro^3 -1} \left(\dfrac{1}{2} \dfrac{\dro^3}{\dr^3} + 1\right) + b \dfrac{1}{1-\nu}\dfrac{1}{\dro-1}\left(\dfrac{1}{2} \dfrac{\dro}{\dr}-\nu\right)\right]
\end{eqnarray}

\subsection{Results for varying Biot's coefficient and Poisson's ratio} \label{sec:elasticResults}
In this section, the results for varying Biot's coefficient and Poisson's ratio are presented.

\begin{figure}
\begin{center}
\includegraphics[width=12.cm]{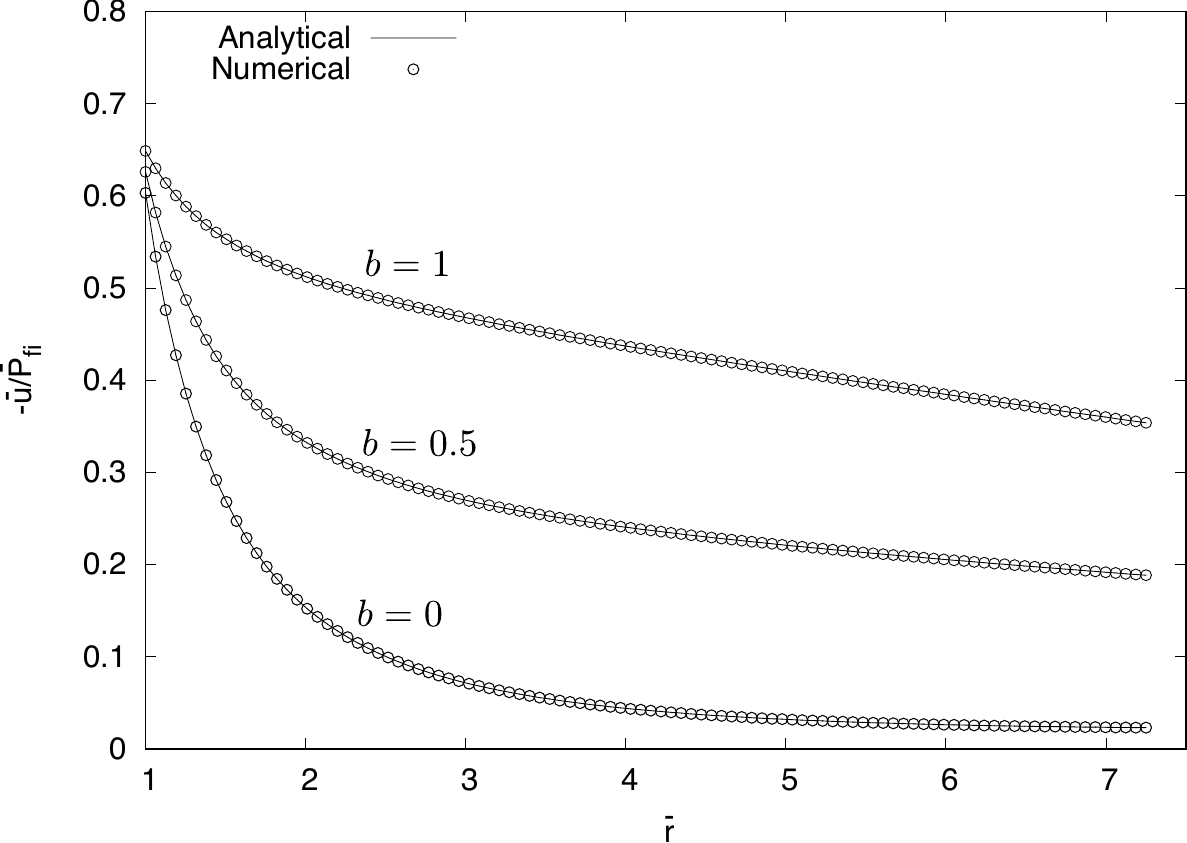}\\
\end{center}
 \caption{Distribution of normalised {\bf radial displacement} plotted as function of dimensionless radial coordinate for Biot's coefficients $b=0$, $0.5$ and $1$, Poisson's ratio $\nu=0.2$, and $\bar{r}_{\rm o} = 7.25$.}
\label{fig:disp}
\end{figure}

\begin{figure}
\begin{center}
\includegraphics[width=12.cm]{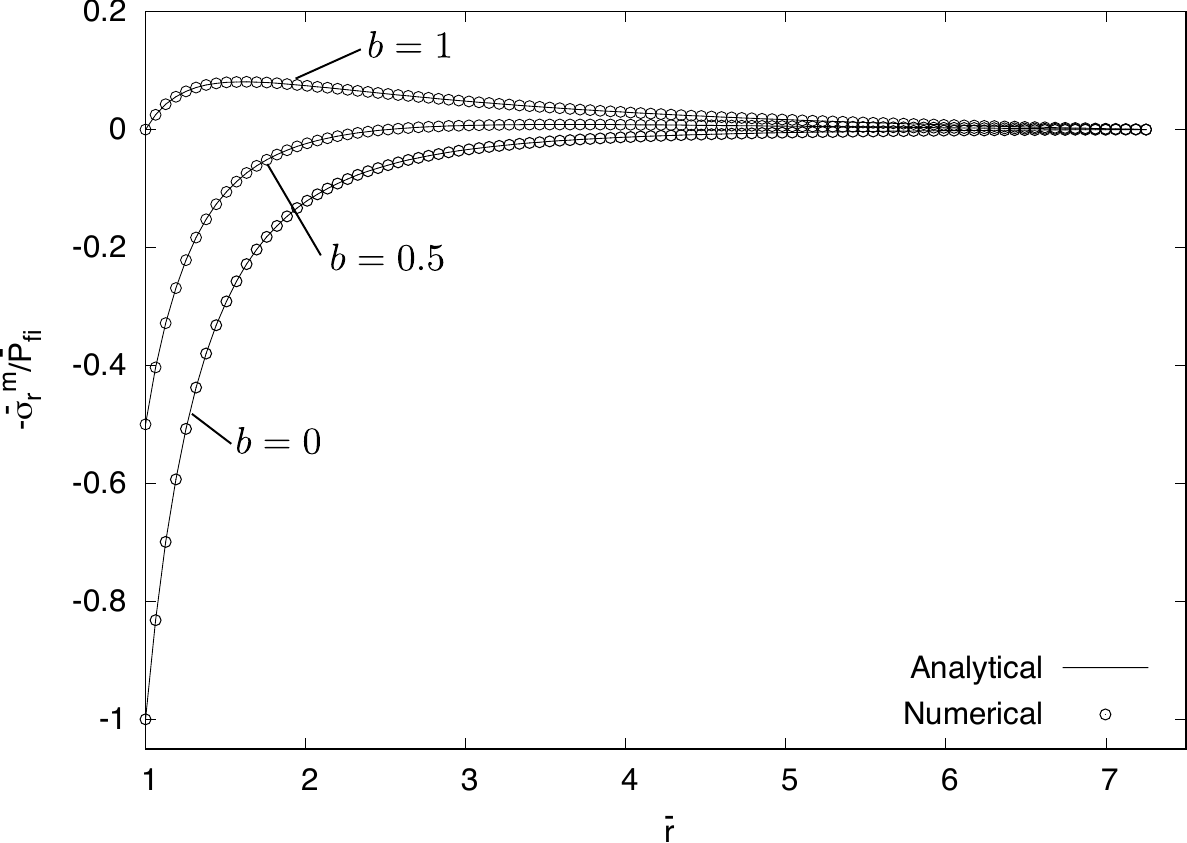}\\
\end{center}
\caption{Distribution of normalised  {\bf effective radial stress} plotted as function of dimensionless radial coordinate for Biot's coefficients $b=0$, $0.5$ and $1$, Poisson's ratio $\nu=0.2$, and $\bar{r}_{\rm o} = 7.25$.}
\label{fig:sigr}
\end{figure}

\begin{figure}
\begin{center}
\includegraphics[width=12.cm]{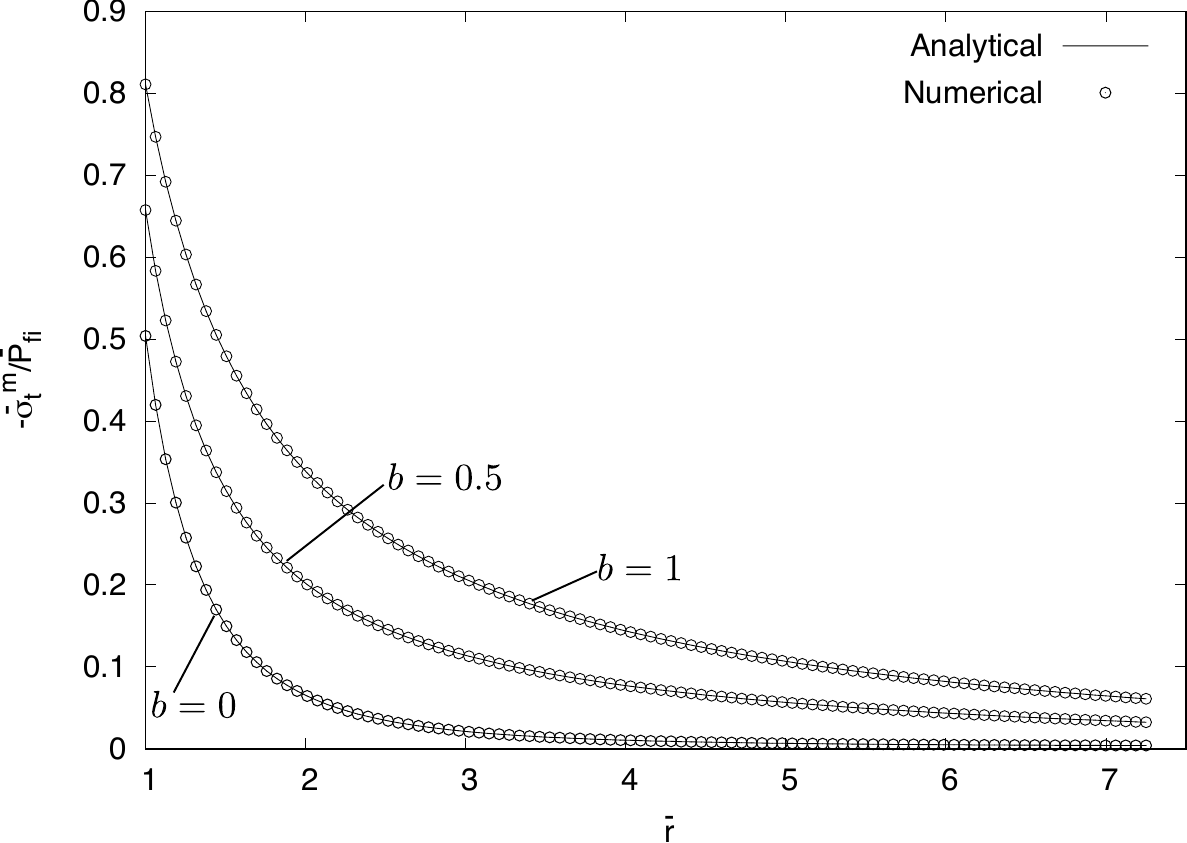}
\end{center}
 \caption{Distribution of normalised {\bf effective tangential stress} plotted as function of dimensionless radial coordinate for Biot's coefficients $b=0$, $0.5$ and $1$, Poisson's ratio $\nu=0.2$, and $\bar{r}_{\rm o} = 7.25$.}
\label{fig:sigt}
\end{figure}

Figures~\ref{fig:disp}--\ref{fig:sigt} show a perfect agreement between the elastic responses calculated by the previous closed-form solution (analytical) and the finite difference code of Appendix~\ref{app:numerical} (numerical). The calculations refer to a Poisson's ratio $\nu=0.2$, a dimensionless outer radius $\bar{r}_{\rm o} = 7.25$ and Biot's coefficients $b=0$, $0.5$ and $1$. The Poisson's ratio was set to 0.2 as this value is representative of most geomaterials and can therefore be used to illustrate a typical elastic response.

In Figures~\ref{fig:disp}--\ref{fig:sigt}, the dimensionless radial displacement, the dimensionless radial stress and the dimensionless tangential stress are normalised by the dimensionless inner pressure changed of sign, $-\dPfi$. Given that the dimensionless fluid pressure $\dPfi$ is compressive (i.e. negative), the minus sign in $-\dPfi$ is necessary to preserve the stress convention of tension positive.

The compressive fluid pressure produces a decrease of the thickness of the spherical wall, which is manifested by a negative difference between the outer and inner radial displacements. The larger is the value of $b$, the smaller is the difference between the two displacements. This means that the most severe compression of the wall of the sphere is obtained for the case of a cavity in a nonporous medium ($b=0$). 

Biot's coefficient has also a strong effect on stresses, with smaller values of $b$ corresponding to larger changes of radial stresses. For $b=0$, the radial stress is always negative (compressive), with the maximum magnitude attained at the inner boundary and a gradual reduction to zero towards the outer boundary. As $b$ increases, the compressive radial stress at the inner boundary becomes smaller while the decay to zero towards the outer boundary is no longer monotonic, which is accompanied by the appearance of tensile radial stresses inside the sphere. For $b=1$, the radial stress is zero at both the inner and outer boundaries with tensile radial stresses at all points inside the sphere.

Finally, the tangential stress is positive for all values of $b$ and attains its maximum value at the inner boundary, with a monotonic decrease towards the outer boundary. Larger values of tangential stress are generated by larger values of $b$.
For all values of $b$, the tangential tensile stress is greater than the radial stress. Therefore, fracture will be initiated at the inner boundary of the thick-walled sphere, as discussed in the next section.

\section{Nonlinear Fracture Response} \label{sec:fracture}

In the present section, the influence of fluid-induced fracture on the response of the thick-walled sphere is investigated.
For the elastic case, it was shown that Biot's coefficient has a strong effect on the mechanical stress.
Here, the influence of this coefficient after the onset of cracking is studied.
In section~\ref{sec:fractureDerivation}, the equations for fluid-induced fracture are derived. Then, the results for varying Biot's coefficient are presented and discussed in section~\ref{sec:fractureBiot}. The influence of size on nominal strength is examined in section~\ref{sec:fractureSize}.

\subsection{Derivation of the equations for fluid-induced fracture}\label{sec:fractureDerivation}
In this section, the equations for fluid-induced fracture are derived.
In a smeared representation, the effect of cracking is reflected by a cracking strain component, which is added to the elastically computed strains.
In the present case, separation of the material is considered to occur only by cracks running in the radial direction, and thus cracking increases the tangential strain only, while the radial strain remains purely elastic.
Formally, this is described by equations
\begin{eqnarray} \label{eq:fracture1}
  \varepsilon_{\rm r} &=& \varepsilon_{\rm r}^{\rm e}
  \\
  \varepsilon_{\rm t} &=& \varepsilon_{\rm t}^{\rm e} + \varepsilon_{\rm t}^{\rm c}
\end{eqnarray}
in which $\varepsilon_{\rm r}^{\rm e}$ and $\varepsilon_{\rm t}^{\rm e}$ are elastic strain components
and $\varepsilon_{\rm t}^{\rm c}$ is the tangential cracking strain.

\begin{figure}
  \begin{center}
    \begin{tabular}{cc}
      \includegraphics[width=6cm]{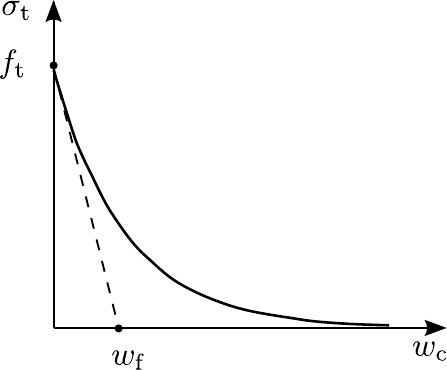} & \includegraphics[width=5.cm]{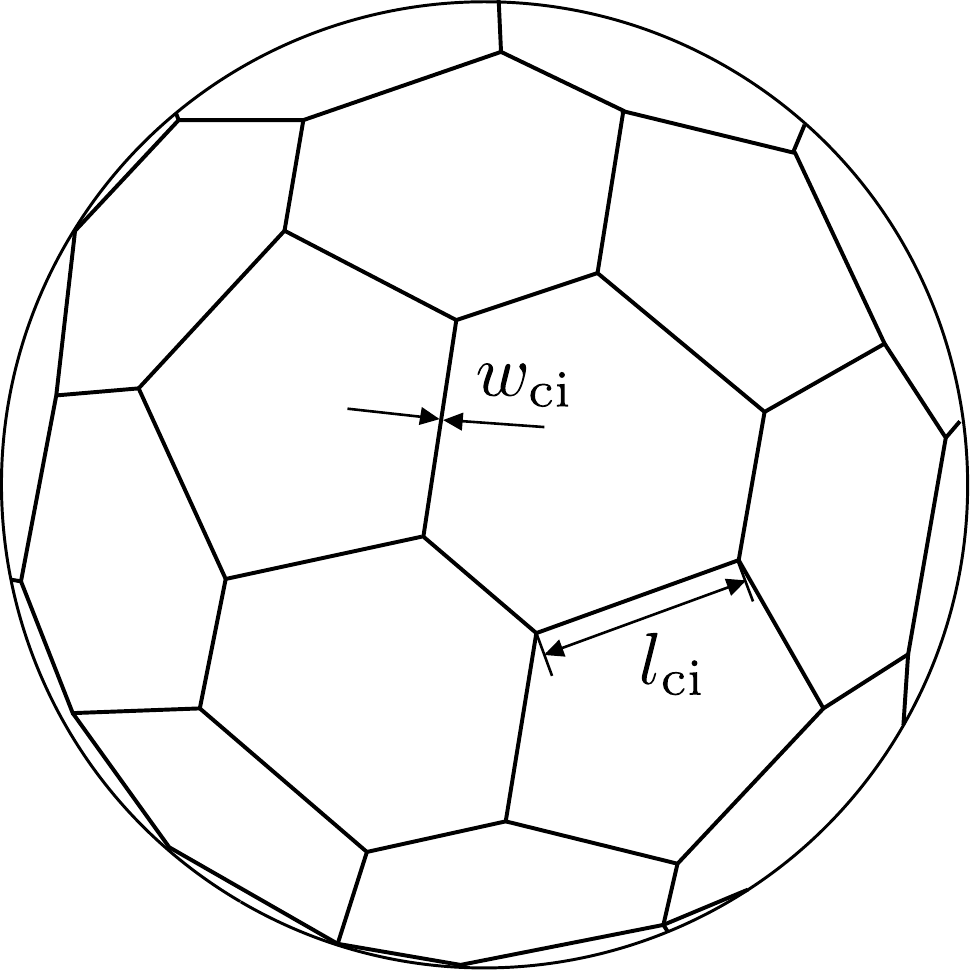}\\
      (a) & (b)
    \end{tabular}
  \end{center}
  \caption{Fracture: (a) exponential stress crack opening curve and (b) possible fracture pattern.}
  \label{fig:softening}
\end{figure}

The elastic stress-strain law (\ref{eq:elastic5x})--(\ref{eq:elastic6x}) remains valid if the tangential strain
is replaced by its elastic part, which can be expressed as $\varepsilon_{\rm t} - \varepsilon_{\rm t}^{\rm c}$.
Combining these modified constitutive equations
\begin{eqnarray} \label{eq:elastic5y}
\sigma_{\rm r}^{\rm m} &=& \dfrac{E}{(1-2\nu)(1+\nu)}\left((1-\nu)\varepsilon_{\rm r}+2\nu(\varepsilon_{\rm t}- \varepsilon_{\rm t}^{\rm c})\right)
\\
\label{eq:elastic6y}
\sigma_{\rm t}^{\rm m} &=& \dfrac{E}{(1-2\nu)(1+\nu)}\left(\nu\varepsilon_{\rm r}+\varepsilon_{\rm t}- \varepsilon_{\rm t}^{\rm c}\right)
\end{eqnarray} 
with kinematic relations (\ref{eq:elastic3})--(\ref{eq:elastic4}) and substituting into equilibrium condition (\ref{eq:elastic2}), we obtain
\begin{equation} \label{eq:fracture3}
   \dfrac{\dd^2 u}{\dd r^2} + 2 \dfrac{\dd u}{\dd r} \dfrac{1}{r} - 2 \dfrac{u}{r^2}-\dfrac{2 \nu}{1-\nu} \dfrac{\dd\varepsilon_{\rm t}^{\rm c}}{\dd r} +\dfrac{2(1-2\nu)}{1-\nu}\dfrac{\varepsilon_{\rm t}^{\rm c}}{r} + b \dfrac{P_{\rm{fi}}}{E} \dfrac{\left(1+\nu\right)\left(1-2\nu\right)}{(1-\nu)}  \dfrac{r_{\rm i} r_{\rm o}}{r_{\rm i}-r_{\rm o}}\dfrac{1}{r^2}=0
\end{equation}
Evolution of the tangential cracking strain $\varepsilon_{\rm t}^{\rm c}$ must be described by a separate law.
In the spirit of traditional smeared crack models \citep{Borst86,Rots88,JirZim98},
it is assumed that $\varepsilon_{\rm t}^{\rm c}$ is
linked to the tangential stress by a softening law, which is postulated here in the exponential form
\begin{equation} \label{eq:fracture4}
\sigma_{\rm t}^{\rm m} =  f_{\rm t} \exp\left(-\dfrac{\varepsilon_{\rm t}^{\rm c}}{\varepsilon_{\rm f}}\right)
\end{equation}
In (\ref{eq:fracture4}), $f_{\rm t}$ is the tensile strength and $\varepsilon_{\rm f}$ is a parameter that controls the steepness of the softening diagram
and is derived from an analogous parameter $w_{\rm f}$ of the exponential stress-crack opening curve shown in Figure~\ref{fig:softening}a.
This curve represents the cohesive response of typical geomaterials (concrete, rocks and stiff soils), which is characterised by an initial steep drop of the cohesive stress followed by a long tail. The area under the stress-crack opening curve is equal to the fracture energy of the material, $G_{\rm F}$. Since the area under the exponential curve is given by the product $f_{\rm t}w_{\rm f}$,
parameter $w_{\rm f}=G_{\rm F}/f_{\rm t}$ can be expressed in terms of physical properties---fracture energy and tensile strength.

Suppose that inelastic deformations localise into a network of cracks that intersect spheres of different radii
in self-similar patterns. An example of such a crack pattern is shown in Figure~\ref{fig:softening}b.
The exact geometry of the pattern is not of importance---what matters is the total length of cracks regularly arranged on a given sphere,
$l_{\rm c}$, which is proportional to the sphere radius, $r$, 
and so we can write
\begin{equation}
  l_{\rm c} = \beta r
\end{equation}
where $\beta$ is a dimensionless parameter characterising the specific crack pattern. Due to the opening $w_{\rm c}$ of localised cracks,
the initial area of the sphere increases by $l_{\rm c}w_{\rm c}$. The effect of cracking can be converted into an equivalent cracking strain
$\varepsilon_{\rm t}^{\rm c}$ uniformly smeared over the sphere, based on the condition that this strain would
lead to the same increase of area. From the corresponding equation
\begin{equation}
  l_{\rm c}w_{\rm c} = 4\pi r^2\times 2\varepsilon_{\rm t}^{\rm c}
\end{equation}
we obtain
\begin{equation}
  \varepsilon_{\rm t}^{\rm c} = \dfrac{l_{\rm c}w_{\rm c}}{8\pi r^2} = \dfrac{\beta rw_{\rm c}}{8\pi r^2}= \dfrac{\beta}{8\pi}\dfrac{w_{\rm c}}{r}
\end{equation}
The same transformation must be applied when a given parameter $w_{\rm f}$ characterising the cohesive crack
is transformed into the corresponding parameter 
\begin{equation}
  \varepsilon_{\rm f} = \dfrac{\beta}{8\pi}\dfrac{w_{\rm f}}{r}
\end{equation}
that is used in the equivalent smeared crack model; see (\ref{eq:fracture4}).
In terms of dimensionless variables, this is rewritten as
\begin{equation}\label{epsf}
  \varepsilon_{\rm f} = \dfrac{\tilde{w}_{\rm f}}{\bar{r}}
\end{equation}
where
\begin{equation}\label{epsf2}
  \tilde{w}_{\rm f}=\dfrac{\beta w_{\rm f}}{8\pi r_{\rm i}}=\dfrac{\beta G_{\rm F}}{8\pi f_{\rm t} r_{\rm i}}
\end{equation}
is a dimensionless parameter that depends on material properties as well as on the inner sphere radius and on the specific crack pattern.

According to (\ref{epsf}), parameter $\varepsilon_{\rm f}$ scales inversely to the radial coordinate. This is a consequence of our assumption that
the inelastic deformations are localised in discrete cracks which intersect concentric surfaces of different radii in a self-similar pattern.
This assumptions seems to be reasonable for the pre-peak regime of the 
fluid-induced fracture process.

The cracking law (\ref{eq:fracture4}) is primarily postulated as a relation between the crack-bridging cohesive stress and the cracking strain. For computational purposes, it is useful to transform the law to a form which links
the cracking strain to the total strain components. This is easily achieved by exploiting
constitutive law (\ref{eq:elastic6y}). Replacing $\sigma_{\rm t}^{\rm m}$ on the left-hand side of (\ref{eq:fracture4}) 
by the expression on the right-hand side of (\ref{eq:elastic6y}), making use of (\ref{epsf})
and rearranging the terms, we construct the equation
\begin{equation} \label{eq:fracture6}
 \varepsilon_{\rm t}^{\rm c} + \left(1+\nu\right)\left(1-2\nu\right)\varepsilon_0 \exp\left(-\frac{\bar{r}\varepsilon_{\rm t}^{\rm c}}{\tilde{w}_{\rm f}}\right) = \varepsilon_{\rm t} +\nu \varepsilon_{\rm r}
\end{equation}
in which $\varepsilon_0=f_{\rm t}/E$ is the limit elastic strain under uniaxial tension.
For given values of total strain components, $\varepsilon_{\rm t}$ and $\varepsilon_{\rm r}$, the corresponding
cracking strain $\varepsilon_{\rm t}^{\rm c}$ is computed by solving nonlinear equation (\ref{eq:fracture6}) iteratively by the Newton method. However, for the sake of generality it is important to mention
that equation (\ref{eq:fracture6}) is valid only during damage growth, i.e., as long as the expression on the right-hand
side is monotonically increasing. Unloading must be treated separately, but since the damage growth is monotonic in
all examples to be presented here, equation (\ref{eq:fracture6}) is fully sufficient for our purpose.
For completeness, possible unloading rules are outlined in Appendix~\ref{app:unload}.

Using the dimensionless variables introduced for the elastic case, (\ref{eq:fracture3}) is transformed into
\begin{equation}\label{eq:fracture15}
\dfrac{\dd^2 \du}{\dd\dr^2} + 2 \dfrac{\dd\du}{\dd\dr} \dfrac{1}{\dr} - 2 \dfrac{\du}{\dr^2}-\dfrac{2 \nu}{1-\nu} \dfrac{\dd\varepsilon_{\rm t}^{\rm c}}{\dd\dr} +\dfrac{2(1-2\nu)}{1-\nu}\dfrac{\varepsilon_{\rm t}^{\rm c}}{\dr} + b \bar{P}_{\rm{fi}} \dfrac{\left(1+\nu\right)\left(1-2\nu\right)}{(1-\nu)}  \dfrac{\dr_{\rm o}}{1- \dr_{\rm o}}\dfrac{1}{\dr^2}=0
\end{equation}
This nonlinear differential equation contains two unknown functions, $\du$ and $\varepsilon_{\rm t}^{\rm c}$,
and it has to be combined with another nonlinear equation (\ref{eq:fracture6}), for the present purpose
rewritten as
\begin{equation} \label{eq:fracture6x}
 \varepsilon_{\rm t}^{\rm c} + \left(1+\nu\right)\left(1-2\nu\right)\varepsilon_0 \exp\left(-\frac{\bar{r}\varepsilon_{\rm t}^{\rm c}}{\tilde{w}_{\rm f}}\right) = \dfrac{\du}{\dr} +\nu \dfrac{\dd\du}{\dd\dr}
\end{equation}
Strictly speaking, equation (\ref{eq:fracture6x}) is applicable only at points that are cracking.
As long as the material remains elastic, equation (\ref{eq:fracture6x}) is replaced by $\varepsilon_{\rm t}^{\rm c}=0$.
The  boundary conditions to be imposed are a slightly modified version of conditions (\ref{eq:elastic100})--(\ref{eq:elastic101}); they read
\begin{eqnarray}\label{eq:elastic100x}
(1-\nu)\dfrac{\dd \du(1)}{\dd \dr}+2\nu\left(\du(1)-\varepsilon_{\rm t}^{\rm c}(1)\right) &=& (1-b) (1-2\nu)(1+\nu) \bar{P}_{\rm fi} 
 \\
 \label{eq:elastic101x}
 (1-\nu)\dfrac{\dd \du(\dr_{\rm o})}{\dd \dr}+2\nu\left(\dfrac{\du(\dr_{\rm o})}{\dr_{\rm o}}-\varepsilon_{\rm t}^{\rm c}(\dr_{\rm o})\right) &=& 0
\end{eqnarray}
The problem is solved numerically using the finite difference method combined with shooting and Newton method.
Details of the numerical procedure are provided in Appendix~\ref{app:numerical}.

The numerically computed global response of the sphere is presented in the form of graphs showing the dependence
between the inner dimensionless fluid pressure and the inner dimensionless radial displacement.
Equilibrium condition written for a half of the sphere implies that the inner pressure times the area of the mid-section of the hole is equal to the integral of the tangential stress over the ligament area, which gives
\begin{equation}\label{eq:fracture17b}
-P_{\rm fi} \pi r_{\rm i}^2 = 2 \int_{r_{\rm i}}^{r_{\rm o}} \sigma_{\rm t} \pi r\, \dd r
\end{equation}
or, in dimensionless form,
\begin{equation}\label{eq:fracture17}
-\bar{P}_{\rm fi} = 2 \int_1^{\bar{r}_{\rm o}} \bar{\sigma}_{\rm t} \bar{r}\, \dd\bar{r}
\end{equation}
The average tangential stress is evaluated as the right hand side of (\ref{eq:fracture17b}) divided by the ligament area,
$\pi \left(r_{\rm o}^2-r_{\rm i}^2\right)$, which results in
\begin{equation}
  \sigma_{\rm t,aver} = \frac{2}{\pi(r_{\rm o}^2-r_{\rm i}^2)} \int_{r_{\rm i}}^{r_{\rm o}} \sigma_{\rm t} \pi r\, \dd r
  =-\frac{P_{\rm fi}\pi r_{\rm i}^2}{\pi(r_{\rm o}^2-r_{\rm i}^2)} = -\frac{P_{\rm fi}}{{r}_{\rm o}^2/r_{\rm i}^2-1}
\end{equation}
or, in dimensionless form, $\bar\sigma_{\rm t,aver} = -\bar{P}_{\rm fi}/(\bar{r}_{\rm o}^2-1)$.
This dimensionless average is used to represent the nonlinear response of the sphere.

\subsection{Results for varying Biot's coefficient}\label{sec:fractureBiot}
\begin{figure}
  \centering
  \begin{tabular}{cc}
 \includegraphics[width=12cm]{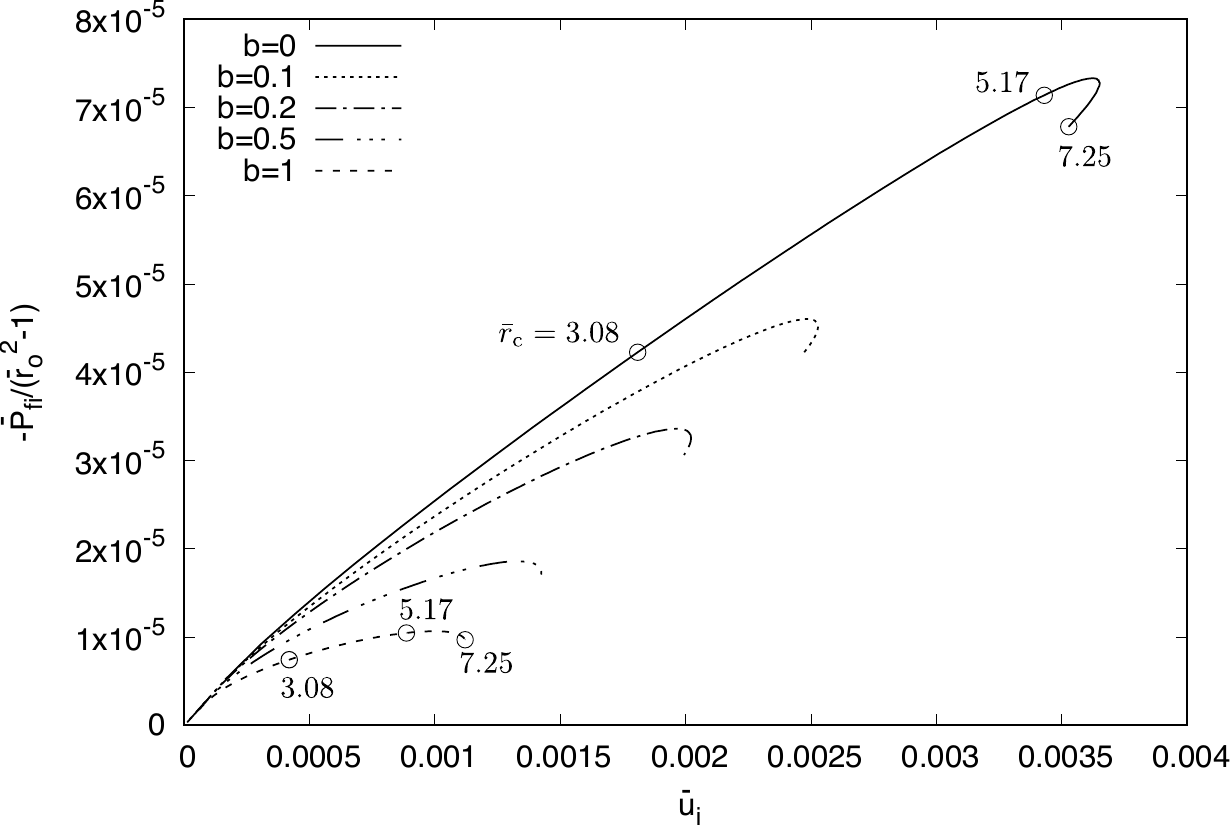}
 \end{tabular}
  \caption{Dimensionless average tangential stress versus dimensionless inner displacement for Poisson's ratio $\nu=0.2$, parameters $\bar{r}_{\rm o} = 7.25$ and $\tilde{w}_{\rm f} = 0.01$, and Biot's coefficients $b = 0$, $0.1$, $0.2$, $0.5$ and $1$.}
  \label{fig:ldBiotNL}
\end{figure}

In this section, the results for varying Biot's coefficient are presented.
Firstly, the dimensionless average tangential stress versus the dimensionless inner displacement is plotted in Figure~\ref{fig:ldBiotNL} for five values of Biot's coefficient ranging from $0$ to $1$.
The individual curves show a strongly nonlinear response, which starts very early in the process. The post-peak response is very brittle, exhibiting a strong snap-back, which is captured in the computation by monotonically increasing the outer displacement $\du(\dr_{\rm o})$ as the control variable.
Biot's coefficient has a strong effect on the average tangential stress.
The highest peak is obtained for $b=0$.
For $b=1$, the peak of the average tangential  stress is less than a fifth of the value for $b=0$, for the specific
values of $\bar{r}_{\rm o}=7.25$ and $\tilde{w}_{\rm f}=0.01$.

The strong effect of $b$ on the average stress is explained by studying the distribution of the tangential stress across the wall of the sphere at three stages of cracking, which are marked in Figure~\ref{fig:ldBiotNL}.
The stages were chosen so that the radial coordinate ${r}_{\rm c}$, which indicates the position of the boundary between the already cracking and yet uncracked parts of the sphere, is equal to $1/3$, $2/3$ and 1 times the ligament
thickness, $r_{\rm o}-r_{\rm i}$.  
The state with $\bar{r}_{\rm c}=1$, i.e., the state at which the outer surface just started cracking,
is located in the post-peak range.
It should be noted that our assumption of self-similar crack patterns only holds for the pre-peak regime.
In the post-peak regime, the inelastic processes can be expected to localise into a few major cracks, which is typical for the propagation stage of hydraulic fracturing.

\begin{figure}
  \centering
 \includegraphics[width=12cm]{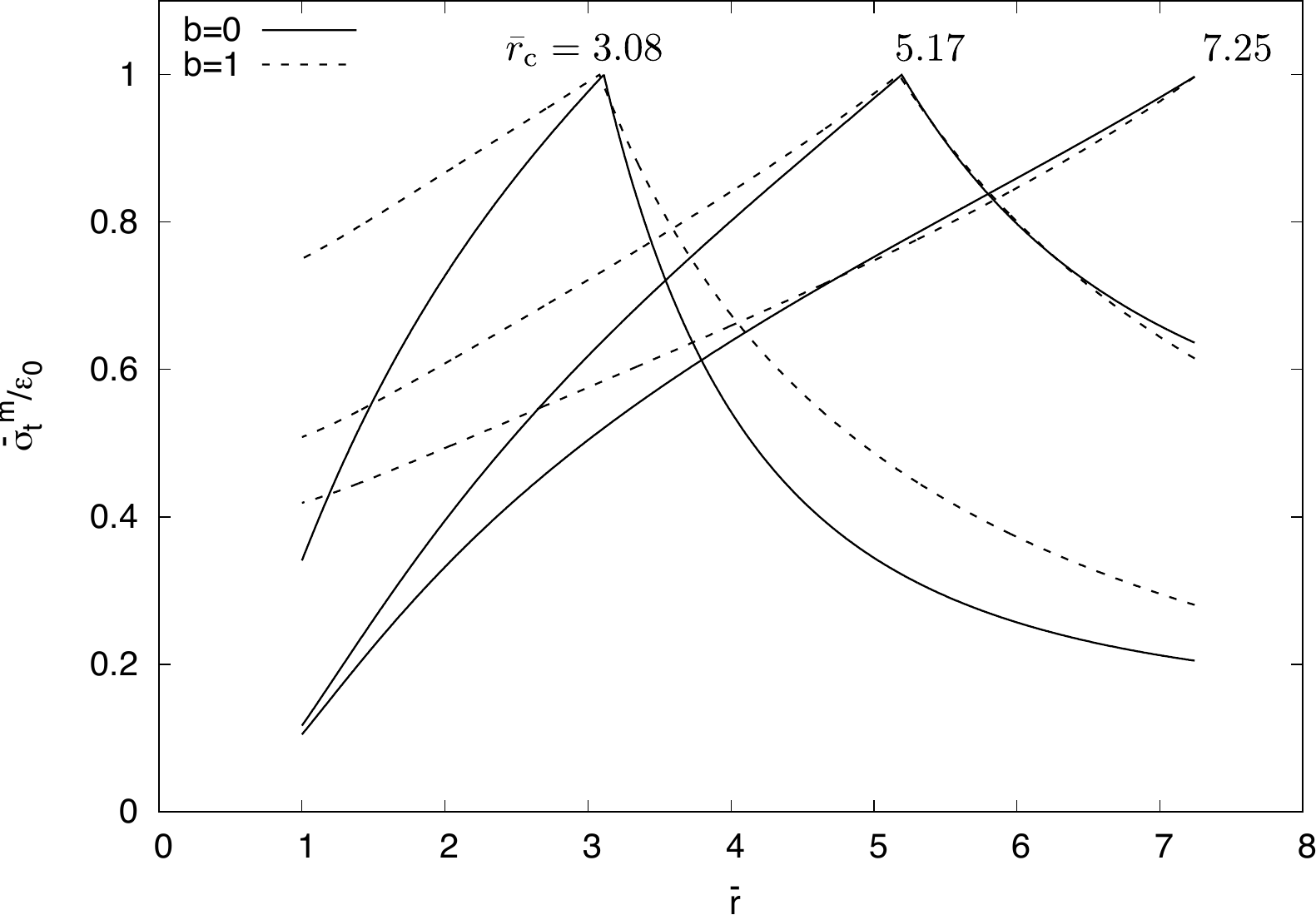}\\
  \caption{Normalised effective tangential stress, $\sigma_{\rm t}^{\rm m}/f_{\rm t}\equiv\bar{\sigma}_{\rm t}^{\rm m}/\varepsilon_0$, versus dimensionless radial coordinate, $\bar{r}$, at three stages marked by hollow circles in Fig.~\ref{fig:ldBiotNL} for $\nu=0.2$ and $b=0$ (solid) or $b=1$ (dashed).}
  \label{fig:sigTNL}
\end{figure}

In Figure~\ref{fig:sigTNL}, the dimensionless effective stress $\bar{\sigma}_{\rm t}^{\rm m}$ divided by the dimensionless tensile strength $\varepsilon_{\rm 0}$ versus the dimensionless radial coordinate $\bar{r}$ is shown for three stages marked in Figure~\ref{fig:ldBiotNL}, with $b=0$ and $1$.
The peaks of the individual curves are equal to the tensile strength and mark the boundary between the cracked and uncracked parts.
For radial coordinates less than the one at which the tensile strength is reached, the material of the sphere undergoes softening.
The rest of the sphere behaves elastically.
The curves for $b=0$ and $b=1$ are similar.

\begin{figure}
  \centering
 \includegraphics[width=12cm]{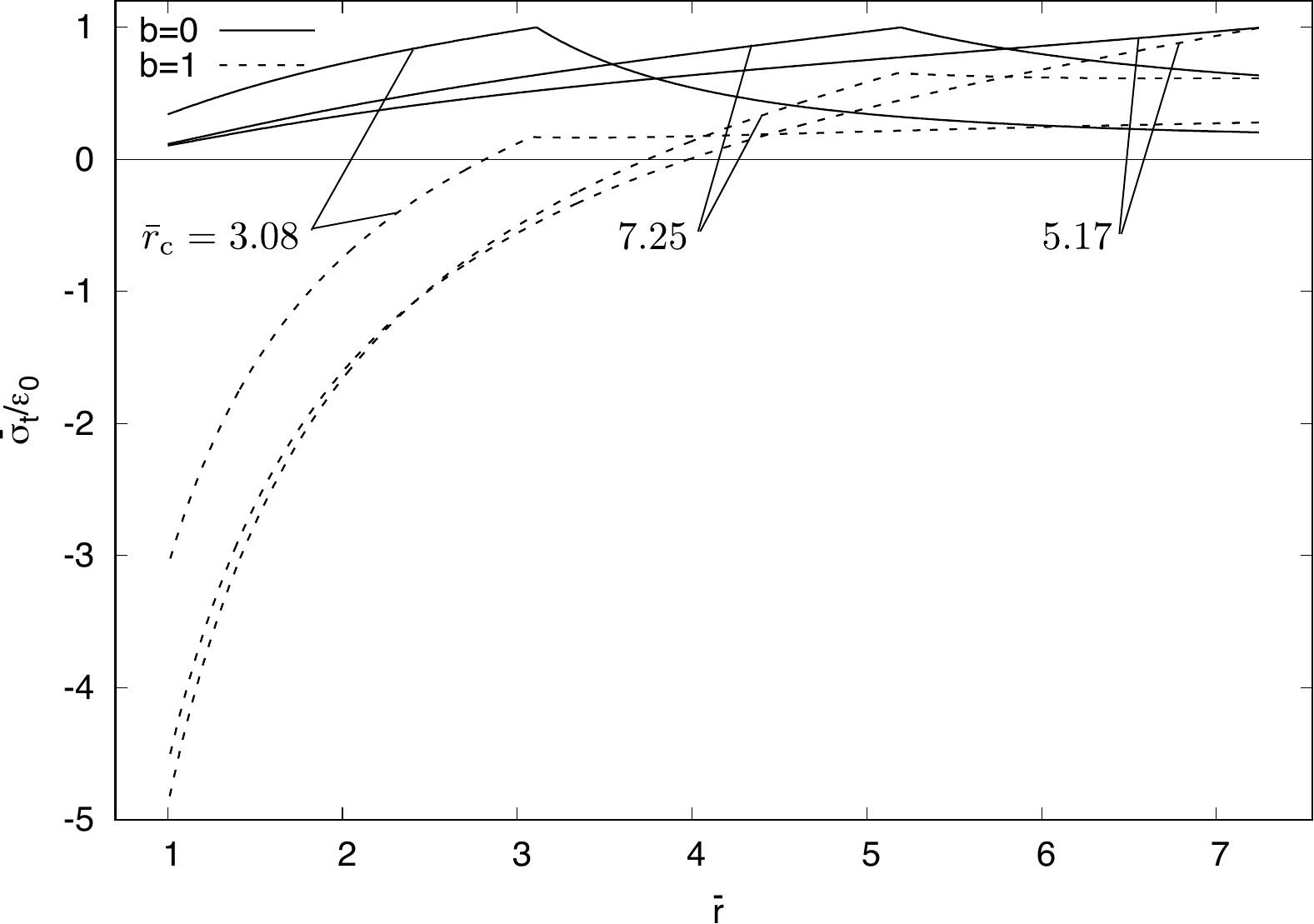}\\
  \caption{Normalised tangential stress, $\sigma_{\rm t}/f_{\rm t}\equiv\bar{\sigma}_{\rm t}/\varepsilon_0$, versus dimensionless radial coordinate, $\bar{r}$, at three stages marked by hollow circles in Fig.~\ref{fig:ldBiotNL} for $\nu=0.2$ and $b=0$ (solid) or $b=1$ (dashed).}
  \label{fig:sigTTotalNL}
\end{figure}

Next, the tangential stress, which enters the equilibrium equation in (\ref{eq:elastic1}), is shown in Figure~\ref{fig:sigTTotalNL} again for the three stages marked in Figure~\ref{fig:ldBiotNL}, with $b=0$ and $1$.
For $b=0$, the total tangential stress is equal to the effective tangential stress shown in Figure~\ref{fig:sigTNL}.
On the other hand, for $b=1$ the tangential stress differs significantly from the one for $b=0$.
At small values of $\bar{r}$, the total tangential stress exhibits negative values of high magnitude, since for $b=1$ the tangential stress is the sum of the effective stress and the fluid pressure. The big difference in the tangential stress distribution explains the strong effect of the fluid pressure on the peak of the average tangential stress.

\subsection{Results for varying sphere size and thickness}\label{sec:fractureSize}
For the present set of results, parameters $\tilde{w}_{\rm f} = 0.01$ and $\bar{r}_{\rm o} = 7.25$ were assumed.
Here, $\tilde{w}_{\rm f} = 0.01$ represents a small sphere. Let us assume a crack length of ten times the circumference, so that $\beta = 20\pi$, a tensile strength of $f_{\rm t} = 3$~MPa and a fracture energy of $G_{\rm F} = 100$~MPa. From (\ref{epsf2}), we can then determine the inner radius as $r_{\rm i} = 8$~mm.
For the effective stress in Figure~\ref{fig:sigTNL}, this results for  $\bar{r}_{\rm c} = \bar{r}_{\rm o}$ in a significant cohesive stress over the entire ligament of the sphere.
The value of the cohesive stress will depend on $\tilde{w}_{\rm f}$ and $\bar{r}_{\rm o}$.
Recall that dimensionless parameter $\tilde{w}_{\rm f}$ is given by (\ref{epsf2}) and depends on the size of the sphere.
For the chosen exponential stress-crack opening law, the characteristic crack opening $w_{\rm f}$ is linked to the fracture energy $G_{\rm F}$ (area under the stress-crack opening curve)
as $w_{\rm f} = {G_{\rm F}}/{f_{\rm t}}$.
Since fracture energy and tensile strength are both material constants, the characteristic crack opening is a material constant as well. Parameter $r_{\rm i}$ represents the size of the sphere, if $\bar{r}_{\rm o}$ is assumed to be constant.
The greater $r_{\rm i}$, the smaller is $\tilde{w}_{\rm f}$.

\begin{figure}
  \centering
  \begin{tabular}{cc}
 \includegraphics[width=5cm]{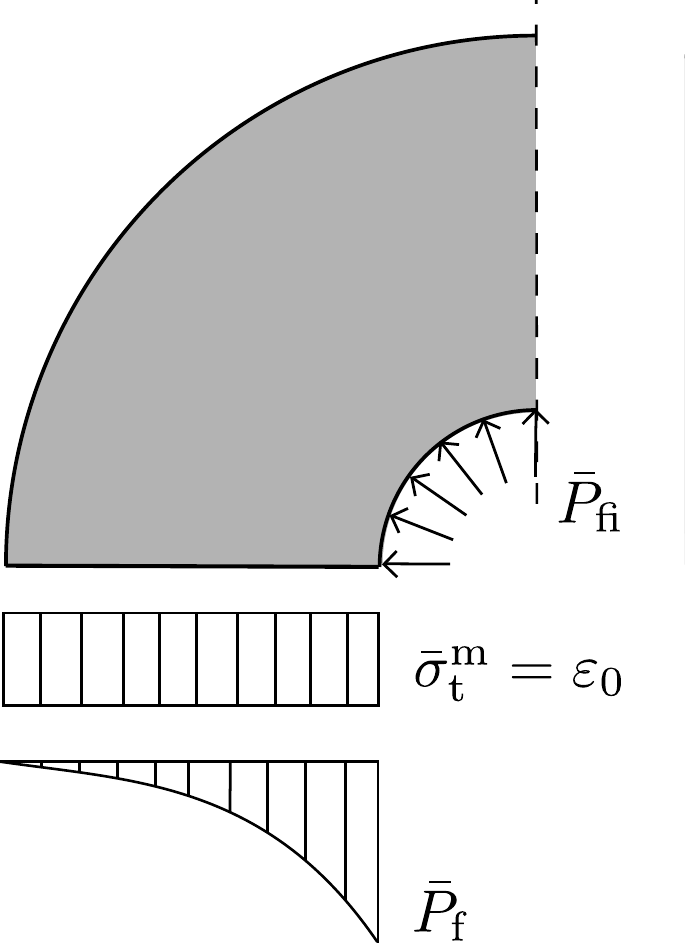} &  \includegraphics[width=5cm]{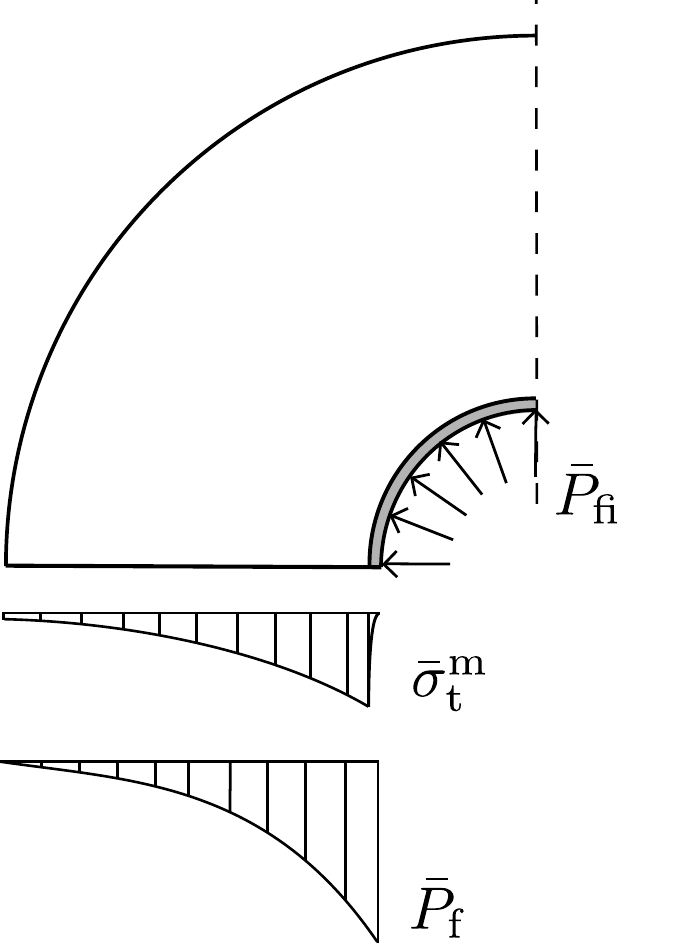}\\
(a) & (b)
  \end{tabular}
  \caption{Schematic illustration of equilibrium for (a) small size ($r_{\rm i} \rightarrow 0$) and (b) large size ($r_{\rm i} \rightarrow \infty$) asymptote.}
  \label{fig:limits}
\end{figure}

\begin{figure}
  \centering
 \includegraphics[width=12cm]{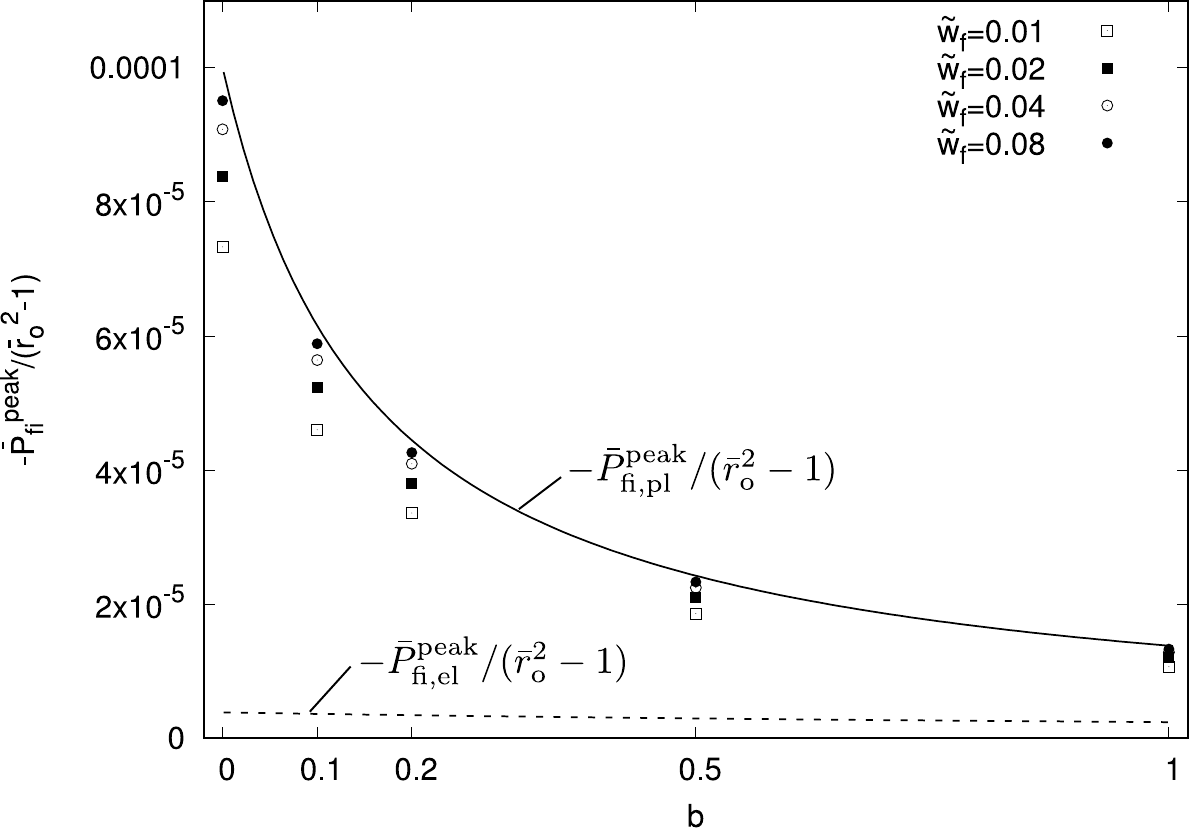}\\
  \caption{Dimensionless strength versus Biot's coefficient for various values of parameter $\tilde{w}_{\rm f}$, with $\dro = 7.25$.}
  \label{fig:sizeEffectRo7p25}
\end{figure}

\begin{figure}
  \centering
 \includegraphics[width=12cm]{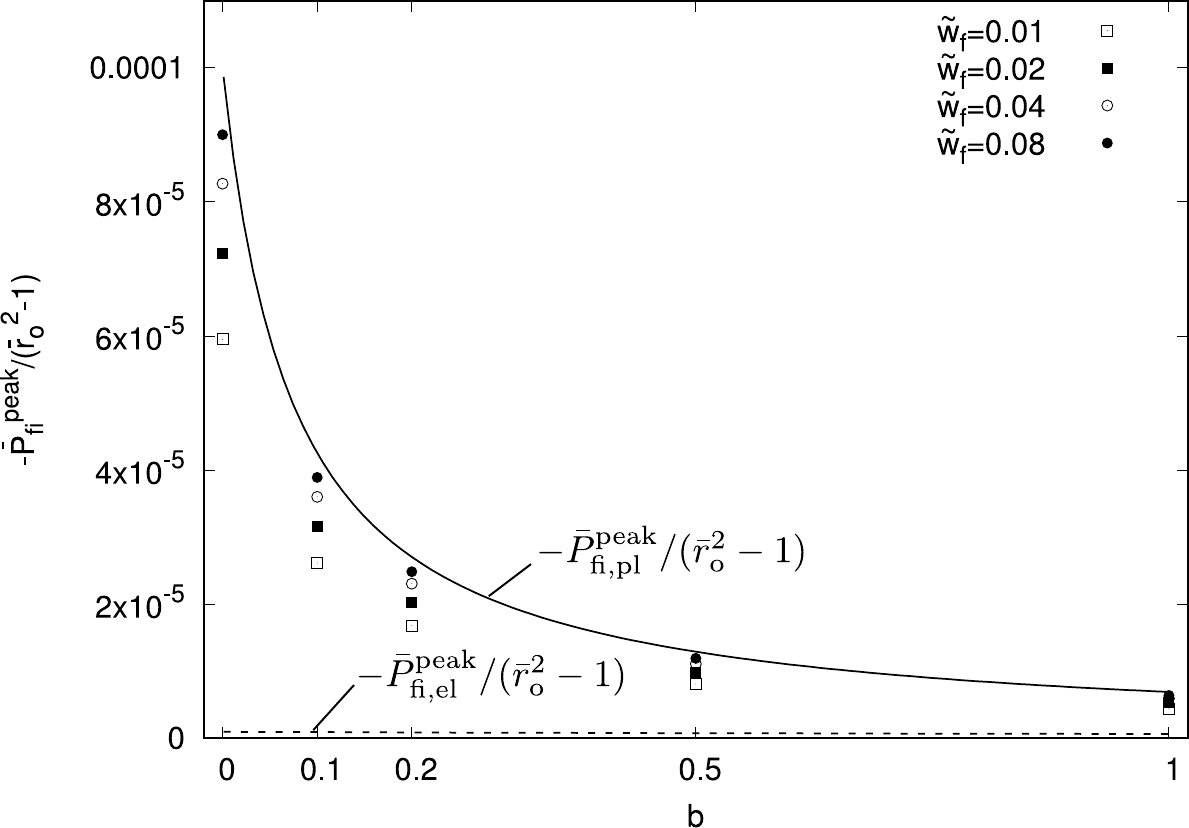}\\
  \caption{Dimensionless strength versus Biot's coefficient for various values of parameter $\tilde{w}_{\rm f}$, with $\dro = 14.5$.}
  \label{fig:sizeEffectRo14p5}
\end{figure}

\begin{figure}
  \centering
 \includegraphics[width=12cm]{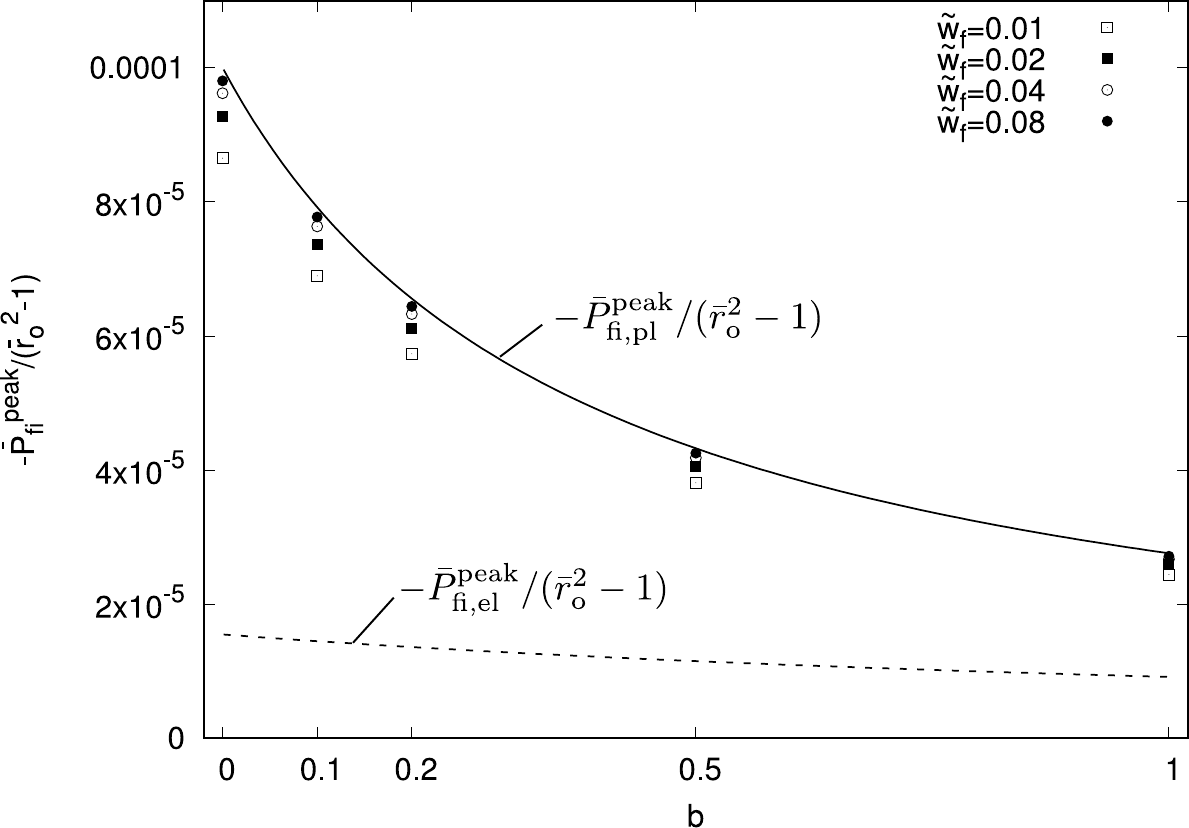}\\
  \caption{Dimensionless strength versus Biot's coefficient for various values of parameter $\tilde{w}_{\rm f}$, with $\dro = 3.125$.}
  \label{fig:sizeEffectRo3p125}
\end{figure}

In the last part of this study, the influence of the size of the sphere on strength, expressed as the peak average tangential stress, is investigated for constant $\bar{r}_{\rm o}$.
Thus, both $r_{\rm i}$ and $r_{\rm o}$ are scaled by the same amount.
The results of the sphere analyses are compared to the small- and large-size asymptotes.
The small-size asymptote for $r_{\rm i} \rightarrow 0$ ($\bar{w}_{\rm fi} \rightarrow \infty$) is derived from a constant distribution of the tangential stress at peak across the ligament area of the thick-walled sphere, as shown in Figure~\ref{fig:limits}a.
The equilibrium equation in (\ref{eq:fracture17}) simplifies  to
\begin{equation}
-\bar{P}_{\rm fi,pl}^{\rm peak} = 2 \int_1^{\bar{r}_{\rm o}} \left(\varepsilon_{0} + b \bar{P}_{\rm f}\right) \, \bar{r} \, \dd \bar{r} = \dfrac{\varepsilon_0}{1+b\left(\bar{r}_{\rm o}-1\right)} \left(\bar{r}_{\rm o}^2 - 1\right)
\end{equation}
Based on (\ref{eq:flow9}), the small size asymptote for the average tangential stress at peak is given by
\begin{equation}
\dfrac{-P_{\rm fi,pl}^{\rm peak}}{\bar{r}_{\rm o}^2 -1} = \dfrac{\varepsilon_{\rm 0}}{1+b\left(\bar{r}_{\rm o} - 1\right)}
\end{equation}
The large-size asymptote corresponds to the case when failure occurs right at the onset of cracking, as shown in Figure~\ref{fig:limits}b. Using the elastic expression of the tangential effective stress, setting it equal to the dimensionless tensile strength ${\varepsilon}_0$ and solving for $\bar{P}_{\rm fi}$ gives
\begin{equation}
  \dfrac{-\bar{P}_{\rm fi,el}^{\rm peak}}{\bar{r}_{\rm o}^2 - 1} =   \dfrac{1}{\bar{r}_{\rm o}^2 - 1} \dfrac{2\varepsilon_0}{\left(1-b \dfrac{1-2\nu}{1-\nu}\right) \dfrac{\dro^3+2}{\dro^3-1} + b \dfrac{1}{1-\nu} \dfrac{\dro-2\nu}{\dro-1}}
\end{equation}
Both of these limits depend strongly on Biot's coefficient, which is one of the main parameters investigated in this study. The maximum average tangential stress $-\bar{P}_{\rm fi}^{\rm peak}$ versus Biot's coefficient for different values of $\tilde{w}_{\rm f}$ are shown in Figures~\ref{fig:sizeEffectRo7p25}, \ref{fig:sizeEffectRo14p5} and \ref{fig:sizeEffectRo3p125} for $\bar{r}_{\rm o} = 7.25$ 14.5 and 3.125, respectively, together with the small- and large-size asymptotes.

There is a strong effect of Biot's coefficient on strength.
The greater Biot's coefficient, the smaller is the strength.
This trend is valid for all sizes, but is most pronounced for small sizes.
Here, the smallest size considered is the one that yields $\tilde{w}_{\rm f} = 0.08$. For this size the strength values are very close to the large-size asymptote. The largest size considered is the one that yields $\tilde{w}_{\rm f} = 0.01$, which was used to produce the results in Figures~\ref{fig:ldBiotNL}~to~\ref{fig:sigTTotalNL}. The strengths obtained for this size are very far from the large-size asymptote. Smaller values of $\tilde{w}_{\rm f}$ could not be considered because of the severity of the snap-back for small Biot's coefficients.

\begin{figure}
  \centering
 \includegraphics[width=12cm]{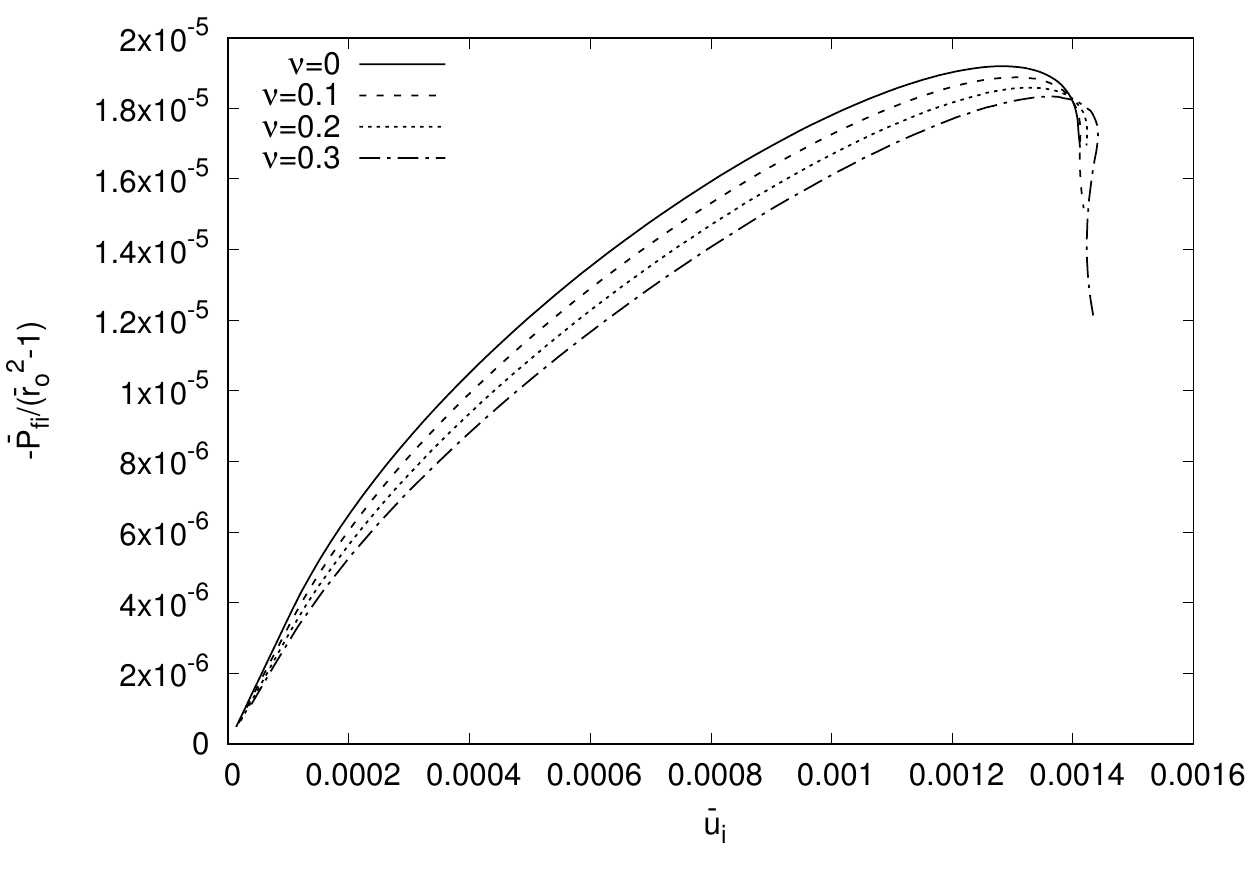}\\
  \caption{Dimensionless average tangential stress versus dimensionless inner displacement for various values of Poisson's ratio $\nu$, with $b=0.5$, $\dro = 7.25$ and $\tilde{w}_{\rm f} =0.01$.}
  \label{fig:ldNu}
\end{figure}

So far, all these nonlinear results have been presented for Poisson's ratio $\nu=0.2$, which was also used for the presentation of the elastic results in Section~\ref{sec:elastic}. In Figure~\ref{fig:ldNu}, the influence of Poisson's ratio is shown to have only a weak influence on the nonlinear response of the sphere. 

\section{Conclusions}
The present study was focused on fracture initiation in a thick-walled hollow sphere made of a permeable material subjected to inner fluid injection.
A new model for fluid-driven fracture initiation taking into account the influence of Biot's coefficient, arbitrary Poisson's ratio and nonlinear fracture mechanics was proposed.
A strong effect of Biot's coefficient on strength for constant sphere geometry was observed.
The greater Biot's coefficient is at constant sphere geometry, the smaller is the nominal strength of the sphere.
Furthermore, the nominal strength depends strongly on the size of the sphere. The greater the size is, the smaller is the strength.
The size effect on nominal strength decreases with increasing Biot's coefficient and decreasing thickness of the sphere.

In future work, it is intended to apply the mathematical model proposed here to the study of physical processes such as damage due to ice formation, salt crystallisation and alkali-silica reactions.
  In the present paper, the geometry of the crack pattern was assumed. It would be interesting to investigate the evolution of the crack pattern from an undamaged state by means of a 3D coupled hydro-mechanical discrete element approach \citep{GraBol16}. With these simulations, the effect of changes of Biot's coefficient due to damage will be investigated as well.

\subsection*{Acknowledgement}
Financial support of the Czech Science Foundation (project GA\v{C}R 17-04150J)
is gratefully acknowledged.

\appendix
    \section{Numerical scheme} \label{app:numerical}

The model developed in Section~\ref{sec:fracture} is mathematically described by two ordinary differential equations
(\ref{eq:fracture15}) and (\ref{eq:fracture6x}), which contain two unknown functions,
$\du$ and $\varepsilon_{\rm t}^{\rm c}$, of a dimensionless variable $\dr$ that ranges from 1 to $\dr_{\rm o}$.
We are interested in the solution that satisfies boundary conditions (\ref{eq:elastic100x})--(\ref{eq:elastic101x}).

For the purpose of numerical implementation, it is useful to
replace the spatial derivative of cracking strain in (\ref{eq:fracture15})
by an equivalent expression in terms of displacement derivatives.
Differentiating (\ref{eq:fracture6x}) with respect to the dimensionless spatial coordinate $\dr$,
we obtain
\begin{equation} \label{eq:fracture61}
 \dfrac{\dd\varepsilon_{\rm t}^{\rm c}}{\dd\dr} - \left(1+\nu\right)\left(1-2\nu\right)\dfrac{\varepsilon_0}{\tilde{w}_{\rm f}} \exp\left(-\frac{\bar{r}\varepsilon_{\rm t}^{\rm c}}{\tilde{w}_{\rm f}}\right)\left(\varepsilon_{\rm t}^{\rm c}+\bar{r}\dfrac{\dd\varepsilon_{\rm t}^{\rm c}}{\dd\dr}\right) = \dfrac{1}{\dr}\dfrac{\dd\du}{\dd\dr} - \dfrac{\du}{\dr^2} +\nu \dfrac{\dd^2\du}{\dd\dr^2}
\end{equation}
To simplify notation, let us introduce an auxiliary parameter
\beq
\phi=\left(1+\nu\right)\left(1-2\nu\right)\dfrac{\varepsilon_0}{\tilde{w}_{\rm f}}
\eeq
and rewrite (\ref{eq:fracture61}) as
\begin{equation} \label{eq:fracture62}
\dfrac{\dd\varepsilon_{\rm t}^{\rm c}}{\dd\dr}\left( 1 - \phi\dr \exp\left(-\frac{\bar{r}\varepsilon_{\rm t}^{\rm c}}{\tilde{w}_{\rm f}}\right)\right) - \varepsilon_{\rm t}^{\rm c}\phi \exp\left(-\frac{\bar{r}\varepsilon_{\rm t}^{\rm c}}{\tilde{w}_{\rm f}}\right) = \dfrac{1}{\dr}\dfrac{\dd\du}{\dd\dr} - \dfrac{\du}{\dr^2} +\nu \dfrac{\dd^2\du}{\dd\dr^2}
\end{equation}
The spatial derivative of cracking strain is now be expressed as
\begin{equation} \label{eq:fracture63}
\dfrac{\dd\varepsilon_{\rm t}^{\rm c}}{\dd\dr} = F(\dr)\left[\dfrac{1}{\dr}\dfrac{\dd\du}{\dd\dr} - \dfrac{\du}{\dr^2} +\nu \dfrac{\dd^2\du}{\dd\dr^2} + \varepsilon_{\rm t}^{\rm c}\phi \exp\left(-\frac{\bar{r}\varepsilon_{\rm t}^{\rm c}}{\tilde{w}_{\rm f}}\right)\right]
\end{equation}
where
\beq\label{eq:F}
F(\dr) = \dfrac{1}{ 1 - \phi\dr \exp\left(-\dfrac{\bar{r}\varepsilon_{\rm t}^{\rm c}(\dr)}{\tilde{w}_{\rm f}}\right)}
\eeq
is an auxiliary function. Finally, substituting (\ref{eq:fracture63}) into
(\ref{eq:fracture15}) and multiplying the whole equation by $1-\nu$, we get
\begin{eqnarray}\nonumber
  (1-\nu-2\nu^2F(\dr))\dfrac{\dd^2 \du}{\dd\dr^2} + 2(1-\nu-\nu F(\dr)) \dfrac{\dd\du}{\dd\dr} \dfrac{1}{\dr} - 2(1-\nu-\nu F(\dr)) \dfrac{\du}{\dr^2}
+ &&
\\
\label{eq:fracture15y}
2(1-2\nu)\dfrac{\varepsilon_{\rm t}^{\rm c}}{\dr}-2 \nu
F(\dr)\phi \exp\left(-\frac{\bar{r}\varepsilon_{\rm t}^{\rm c}}{\tilde{w}_{\rm f}}\right)\varepsilon_{\rm t}^{\rm c} + b \bar{P}_{\rm{fi}} \left(1+\nu\right)\left(1-2\nu\right)  \dfrac{\dr_{\rm o}}{1- \dr_{\rm o}}\dfrac{1}{\dr^2}&=&0
\end{eqnarray}
To extend the validity of this equation to the regions which have
not started cracking yet, it is sufficient to set $F(\dr)=0$ for all $\dr$ at which $\varepsilon_{\rm t}^{\rm c}(\dr)=0$. Therefore, the precise definition of function $F$ is
\beq\label{eq:FF}
F(\dr) = \left\{\begin{array}{ll}
\dfrac{1}{ 1 - \phi\dr \exp\left(-\bar{r}\varepsilon_{\rm t}^{\rm c}(\dr)/\tilde{w}_{\rm f}\right)}
& \mbox{ if } \varepsilon_{\rm t}^{\rm c}(\dr)>0
\\
0 & \mbox{ if } \varepsilon_{\rm t}^{\rm c}(\dr)=0
\end{array}\right.
\eeq

The numerical procedure is based on replacement of spatial derivatives
in equation (\ref{eq:fracture15y}) by finite differences.
Recall that we are interested in the solution
that satisfies boundary conditions (\ref{eq:elastic100x})--(\ref{eq:elastic101x}).
One of these conditions is imposed at $\dr=1$ and the other at $\dr=\dr_{\rm o}$.
To avoid the need for solving a large set of discretised algebraic equations, we use the
shooting method, which converts the boundary value problem to an initial value problem.
The main idea is that, at one boundary point, the true physical boundary condition is supplemented
by another, fictitious boundary condition, and then the numerical solution can be computed over the
whole interval in an explicit way. Of course, for an arbitrary choice of the fictitious boundary condition,
the true physical boundary condition at the other end of the interval is in general not satisfied.
Therefore, the value prescribed by the fictitious boundary condition is iterated until the
boundary condition at the other end is satisfied. This can be considered as the solution of one
nonlinear equation, which can be performed, e.g., by the Newton method.

The approach described above could be applied in a straightforward manner 
if the loading process is controlled by increasing the applied inner pressure, $\bar{P}_{\rm fi}$.
However, this would work only in the pre-peak range of the load-displacement diagram and the
post-peak response could not be captured. Due to the highly brittle post-peak behaviour,
direct displacement control with prescribed displacement at the inner boundary would fail
shortly after the peak since typical load-displacement diagrams exhibit snapback. It turns
out that a suitable control variable is the displacement at the outer boundary, which
can be monotonically increased under indirect displacement control. This results into
a modified version of the shooting method, in which the additional boundary condition
imposed on the outer boundary is actually fixed, based on the prescribed value of the control
variable, and the variable on which we iterate is the inner pressure. Consequently,
the integration process starts at the outer boundary and proceeds ``backwards'' to the
inner boundary. In each global increment, the displacement $\bar{u}_{\rm o}$ at the outer
boundary kept fixed, and the objective of the shooting method is to find the
inner pressure $\bar{P}_{\rm fi}$ for which the numerically computed solution satisfies
boundary condition (\ref{eq:elastic100x}) on the inner boundary.

In order to construct a numerical solution, the interval $[1,\dr_{\rm o}]$ is divided into $N$ equal subintervals of length $h=(\dr_{\rm o}-1)/N$,
separated by grid points $\dr_k=1+kh$, $k=0,1,\ldots N$, and we search for
approximations of displacements and cracking
strains at the grid points denoted as $\du_k$ and $\varepsilon_{{\rm t},k}^{\rm c}$.
The integration scheme is initialised by imposing two
conditions on the outer boundary, i.e., at $\dr=\dr_{\rm o}\equiv\dr_N$. One of these conditions,
\beq\label{ee115a}
\du_N=\bar{u}_{\rm o}
\eeq
has just been explained, and the other is simply the true physical
boundary condition (\ref{eq:elastic101x}), in the discretised form rewritten as
\beq\label{ee115b}
 (1-\nu)\dfrac{\du_{N+1}-\du_{N-1}}{2h}+2\nu\left(\dfrac{\bar{u}_{\rm o}}{\dr_N}-\varepsilon_{{\rm t},N}^{\rm c}\right) = 0
\eeq
from which it is easy to express
\beq\label{ee115c}
\du_{N+1}=\du_{N-1}-\dfrac{4\nu h}{1-\nu}\left(\dfrac{\bar{u}_{\rm o}}{\dr_N}-\varepsilon_{{\rm t},N}^{\rm c}\right)
 \eeq
However, note that the resulting expression contains the cracking strain at the outer boundary,
$\varepsilon_{{\rm t},N}^{\rm c}$, which is not a priori known.

One can first assume that the
material remains in an elastic state, in which case $\varepsilon_{{\rm t},N}^{\rm c}=0$.
 This elastic trial solution is admissible only if the corresponding elastically evaluated
effective tangential stress does not exceed the tensile strength, which is in the dimensionless
form written as
\beq\label{ee115d}
\nu\dfrac{\du_{N+1}-\du_{N-1}}{2h}+\dfrac{\bar{u}_{\rm o}}{\dr_N} \le (1-2\nu)(1+\nu)\varepsilon_0
\eeq
Substituting from (\ref{ee115c}) with  $\varepsilon_{{\rm t},N}^{\rm c}$ set to zero,
one can show that condition (\ref{ee115d}) is equivalent to
$\bar{u}_{\rm o}\le (1-\nu)\varepsilon_0\dr_N $, which can be readily checked before
the evaluation of (\ref{ee115c}).

If the prescribed displacement $\bar{u}_{\rm o}$
exceeds the limit value $(1-\nu)\varepsilon_0\dr_N$, then the material on the right boundary is cracking
and equation (\ref{ee115b}) needs to be combined with equation (\ref{eq:fracture6x}), written
at $\dr=\dr_N$ in the discretised form
\begin{equation} \label{eq:fracture66N}
 \varepsilon_{{\rm t},N}^{\rm c} + \phi\tilde{w}_{\rm f} \exp\left(-\frac{\bar{r}_N\varepsilon_{{\rm t},N}^{\rm c}}{\tilde{w}_{\rm f}}\right) = \dfrac{\du_N}{\dr_N} +\nu \dfrac{\du_{N+1}-\du_{N-1}}{2h}
\end{equation}
Based on (\ref{ee115a}) and (\ref{ee115c}), the right-hand side of (\ref{eq:fracture66N})
can be expressed in terms of known quantities and $\varepsilon_{{\rm t},N}^{\rm c}$ as the only unknown, and the
resulting equation
\begin{equation} \label{eq:fracture66Nx}
 \varepsilon_{{\rm t},N}^{\rm c} + (1-\nu)\varepsilon_0\exp\left(-\frac{\bar{r}_N\varepsilon_{{\rm t},N}^{\rm c}}{\tilde{w}_{\rm f}}\right) = \dfrac{\bar{u}_{\rm o}}{\dr_N} 
\end{equation}
can be solved by the Newton method, starting from the initial guess $\varepsilon_{{\rm t},N}^{\rm c}=0$.
Afterwards, $\du_{N+1}$
is evaluated from (\ref{ee115c}), which makes it possible to start the regular stepping procedure
from the outer boundary,  because the values of $\du_N$, $\du_{N+1}$ and $\varepsilon_{{\rm t},N}^{\rm c}$ are now known.

In a generic step $k$ (with $k$ decreasing from $N$ to 1),
the values of $\du_k$, $\du_{k+1}$ and $\varepsilon_{{\rm t},k}^{\rm c}$ are known,
and the values of $\du_{k-1}$ and $\varepsilon_{{\rm t},k-1}^{\rm c}$ need to be computed.
At point
$\dr=\dr_k$, equation (\ref{eq:fracture15y}) is approximated by
\begin{eqnarray}\nonumber
  (1-\nu-2\nu^2F_k)\frac{\du_{k+1}-2\du_k+\du_{k-1}}{h^2} + 2(1-\nu-\nu F_k)\frac{\du_{k+1}-\du_{k-1}}{2hr_k}  - 2(1-\nu-\nu F_k) \frac{\du_k}{r_k^2}
+ &&
\\
\label{eq:fracture15yz}
2(1-2\nu)\dfrac{\varepsilon_{{\rm t},k}^{\rm c}}{\dr_k}-2 \nu
F_k\phi \exp\left(-\frac{\bar{r}_k\varepsilon_{{\rm t},k}^{\rm c}}{\tilde{w}_{\rm f}}\right)\varepsilon_{{\rm t},k}^{\rm c} + b \bar{P}_{\rm{fi}} \left(1+\nu\right)\left(1-2\nu\right)  \dfrac{\dr_{\rm o}}{1- \dr_{\rm o}}\dfrac{1}{\dr_k^2}&=&0
\end{eqnarray}
in which $F_k=F(\dr_k)$ is the numerical values of function $F$ defined in (\ref{eq:FF}) at $\dr=\dr_k$.
Equation (\ref{eq:fracture15yz}) can be rewritten as
\beq\label{ee108}
A_k\frac{\du_{k+1}-2\du_k+\du_{k-1}}{h^2} +B_k\frac{\du_{k+1}-\du_{k-1}}{2h} +C_k \du_k+
D_k=0
\eeq
where
\begin{eqnarray}
A_k &=& 1-\nu-2 \nu^2F_k
\\
B_k &=& \dfrac{2\left(1-\nu-\nu F_k\right)}{\dr_k}
\\
C_k &=& -\dfrac{2\left(1-\nu-\nu F_k\right)}{\dr_k^2}
\\
\label{ee109d}
D_k &=& 2(1-2\nu)\dfrac{\varepsilon_{{\rm t},k}^{\rm c}}{\dr_k}-2 \nu
F_k\phi \exp\left(-\frac{\bar{r}_k\varepsilon_{{\rm t},k}^{\rm c}}{\tilde{w}_{\rm f}}\right)\varepsilon_{{\rm t},k}^{\rm c} + b \bar{P}_{\rm{fi}} \left(1+\nu\right)\left(1-2\nu\right)  \dfrac{\dr_{\rm o}}{1- \dr_{\rm o}}\dfrac{1}{\dr_k^2}
\end{eqnarray}
Using these auxiliary coefficients, the displacement update formula
derived from (\ref{ee108}) can be conveniently written as
\beq\label{ee113}
\du_{k-1} = \dfrac{\left(\dfrac{A_k}{h^2}+\dfrac{B_k}{2h}\right)\du_{k+1}+\left(C_k-\dfrac{2A_k}{h^2}\right)\du_{k}+D_k}{\dfrac{B_k}{2h}-\dfrac{A_k}{h^2}}
\eeq
To finish the step, it is also necessary to evaluate the cracking strain $\varepsilon_{{\rm t},k-1}$.
This is done using equation (\ref{eq:fracture6x}) written  at $\dr_{k-1}$ as
\begin{equation} \label{eq:fracture67}
 \varepsilon_{{\rm t},k-1}^{\rm c} + \phi\tilde{w}_{\rm f} \exp\left(-\frac{\bar{r}_{k-1}\varepsilon_{{\rm t},k-1}^{\rm c}}{\tilde{w}_{\rm f}}\right) = \dfrac{\du_{k-1}}{\dr_{k-1}} +\nu \du'_{k-1}
\end{equation}
where $\du'_{k-1}$ is a suitable approximation of $\dd\du/\dd\dr$ at $\dr=\dr_{k-1}$.
Normally, this approximation would be provided by the central difference expression,
$(\du_k-\du_{k-2})/2h$,
but since the value of $\du_{k-2}$
is not known yet, an alternative second-order accurate approximation
\beq\label{eq:fracture68}
\du'_{k-1} =\dfrac{\du_{k+1}-\du_{k-1}}{2h}-h\dfrac{\du_{k+1}-2\du_k+\du_{k-1}}{h^2} = \dfrac{-\du_{k+1}+4\du_k-3\du_{k-1}}{2h}
\eeq
is constructed based on the already known displacement values. Making use of (\ref{eq:fracture68}),
the right-hand side of (\ref{eq:fracture67}) is easily evaluated. If the result is smaller than
$\phi\tilde{w}_{\rm f}$ (which is equal to $(1-\nu-2\nu^2)\varepsilon_0$), then there is no cracking and $\varepsilon_{{\rm t},k-1}$ is set to zero,
otherwise $\varepsilon_{{\rm t},k-1}$ is computed from equation  (\ref{eq:fracture67}) by the Newton method,
using $\varepsilon_{{\rm t},k}$ as the initial estimate.

The generic step is repeated until counter $k$ becomes equal to 1, which means that the integration
process reaches the inner boundary. The computed approximate solution is admissible only if it satisfies
boundary condition (\ref{eq:elastic100x}), which is in terms of the numerical values written as
\beq\label{eq:elastic100y}
(1-\nu)\dfrac{-\du_{2}+4\du_1-3\du_{0}}{2h}+2\nu\left(\du_0-\varepsilon_{{\rm t},0}^{\rm c}\right) = (1-b) (1-2\nu)(1+\nu) \bar{P}_{\rm fi}
\eeq
The difference between the right-hand side and the left-hand side of (\ref{eq:elastic100y}) is the residuum
of the shooting method, considered as a function of the inner pressure, $\bar{P}_{\rm fi}$.
Of course, this function is in general nonlinear, because $\bar{P}_{\rm fi}$ affects not only the right-hand side,
where it appears explicitly, but also the left-hand side, since it has an influence on coefficients
$D_k$ computed according to (\ref{ee109d}) and thus on the entire numerical solution, including the resulting values
of displacements and cracking strain that appear in (\ref{eq:elastic100y}).
In the top loop of the shooting method,
the value of $\bar{P}_{\rm fi}$ is iteratively adjusted and the numerical solution is recomputed
until the residual becomes negligible.

\section{Extension of cracking law to unloading}
\label{app:unload}

In the examples treated in this paper, the cracking process at each material point is monotonic,
and so the straightforward description of the cohesive law by equation (\ref{eq:fracture4}) is sufficient.
It would not be difficult to extend this description to the general case with possible unloading.
A frequently used simple assumption is that, during unloading, cracks are closing and the cracking
strain decreases in proportion to the normal stress transmitted by the cohesive crack. 
The corresponding stress-cracking strain law can be written as
\begin{equation}\label{eq:fracture7}
\sigma_{\rm t}^{\rm m} = C \varepsilon_{\rm t}^{\rm c}
\end{equation}
where 
\begin{equation}\label{eq:fracture8}
C = \frac{f_{\rm t}}{\varepsilon_{\rm t}^{\rm{c,max}}}\exp\left(-\frac{\varepsilon_{\rm t}^{\rm{c,max}}}{\varepsilon_{\rm f}}\right)
\end{equation}
is the unloading stiffness, dependent on
the maximum previously reached value of cracking strain, $\varepsilon_{\rm t}^{\rm{c,max}}$.

Substituting (\ref{eq:fracture7}) into the stress-strain equation (\ref{eq:elastic6y}), which still remains valid,
we obtain
\begin{equation}\label{eq:fracture7f}
C \varepsilon_{\rm t}^{\rm c} = \frac{E}{(1-2\nu)(1+\nu)}\left(\nu\varepsilon_{\rm r} + \varepsilon_{\rm t} - \varepsilon_{\rm t}^{\rm c}\right)
\end{equation}
This is a linear equation
from which the dependence of the cracking strain on total strains during unloading can be determined:
\begin{equation}\label{eq:fracture10}
\varepsilon_{\rm t}^{\rm c}  =  \dfrac{\varepsilon_{\rm t} + \nu \varepsilon_{\rm r}}{1+\dfrac{1-\nu-2\nu^2}{E}C}
\end{equation}
Subsequently, the cracking strain can be eliminated from (\ref{eq:elastic5y})--(\ref{eq:elastic6y}) and the stress-strain equations can be written in the form of an elastic law with reduced stiffness coefficients,
\begin{eqnarray}\label{eq:fracture11}
\sigma_{\rm r} &=& \dfrac{E}{1-\nu-2\nu^2} \left(\left(1-\nu-\frac{2\nu^2 E}{E+C}\right)\varepsilon_{\rm r} +\frac{2\nu C}{E+C}\varepsilon_{\rm t}\right)
\\
\label{eq:fracture12}
\sigma_{\rm t} &=& \dfrac{E}{1-\nu-2\nu^2} \left(\frac{\nu C}{E+C}\varepsilon_{\rm r} +\frac{C}{E+C}\varepsilon_{\rm t}\right)
\end{eqnarray}
Note that symmetry is preserved, because $\sigma_{\rm t}$ is work-conjugate with $2\varepsilon_{\rm t}$.
In matrix form, equations (\ref{eq:fracture11})--(\ref{eq:fracture12}) could be written as
\begin{equation}\label{eq:fracture13}
\left(\begin{array}{c}\sigma_{\rm r} \\ \sigma_{\rm t}\end{array}\right)
=
 \dfrac{E}{1-\nu-2\nu^2} \left(\begin{array}{cc}1-\nu-\dfrac{2\nu^2 E}{E+C} &  \dfrac{\nu C}{E+C}\\[2mm] \dfrac{\nu C}{E+C} & \dfrac{C}{2(E+C)} \end{array}\right)
\left(\begin{array}{c}\varepsilon_{\rm r} \\ 2\varepsilon_{\rm t}\end{array}\right)
\end{equation}
It is also interesting to note that, for a fully formed stress-free crack characterised by $C=0$,
equation (\ref{eq:fracture13}) reduces to
\begin{equation}\label{eq:fracture14}
\left(\begin{array}{c}\sigma_{\rm r} \\ \sigma_{\rm t}\end{array}\right)
=
\left(\begin{array}{cc}E &  0\\[2mm] 0 & 0 \end{array}\right)
\left(\begin{array}{c}\varepsilon_{\rm r} \\ 2\varepsilon_{\rm t}\end{array}\right)
\end{equation}
In this case, the stress $\sigma_{\rm t}$ transmitted by the crack vanishes, and the stress $\sigma_{\rm r}$ parallel to the crack is linked to the radial strain by the simple form of Hooke's law, valid for uniaxial stress.

\bibliographystyle{elsarticle-harv}
\bibliography{general}

\begin{thebibliography}{46}
\expandafter\ifx\csname natexlab\endcsname\relax\def\natexlab#1{#1}\fi
\providecommand{\url}[1]{\texttt{#1}}
\providecommand{\href}[2]{#2}
\providecommand{\path}[1]{#1}
\providecommand{\DOIprefix}{doi:}
\providecommand{\ArXivprefix}{arXiv:}
\providecommand{\URLprefix}{URL: }
\providecommand{\Pubmedprefix}{pmid:}
\providecommand{\doi}[1]{\href{http://dx.doi.org/#1}{\path{#1}}}
\providecommand{\Pubmed}[1]{\href{pmid:#1}{\path{#1}}}
\providecommand{\bibinfo}[2]{#2}
\ifx\xfnm\relax \def\xfnm[#1]{\unskip,\space#1}\fi
\bibitem[{Adachi et~al.(2007)Adachi, Siebrits, Peirce and
  Desroches}]{AdaSiePei07}
\bibinfo{author}{Adachi, J.}, \bibinfo{author}{Siebrits, E.},
  \bibinfo{author}{Peirce, A.}, \bibinfo{author}{Desroches, J.},
  \bibinfo{year}{2007}.
\newblock \bibinfo{title}{Computer simulation of hydraulic fractures}.
\newblock \bibinfo{journal}{International Journal of Rock Mechanics and Mining
  Sciences} \bibinfo{volume}{44}, \bibinfo{pages}{739--757}.
\bibitem[{Ahmed et~al.(2017)Ahmed, Siddique and Mahmood}]{AhmSidMah17}
\bibinfo{author}{Ahmed, A.}, \bibinfo{author}{Siddique, J.I.},
  \bibinfo{author}{Mahmood, A.}, \bibinfo{year}{2017}.
\newblock \bibinfo{title}{Non-newtonian flow-induced deformation from
  pressurized cavities in absorbing porous tissues}.
\newblock \bibinfo{journal}{Computer Methods in Biomechanics and Biomedical
  Engineering} \bibinfo{volume}{20}, \bibinfo{pages}{1464--1473}.
\bibitem[{Andrade et~al.(1993)Andrade, Alonso and Molina}]{AndAloMol93}
\bibinfo{author}{Andrade, C.}, \bibinfo{author}{Alonso, C.},
  \bibinfo{author}{Molina, F.J.}, \bibinfo{year}{1993}.
\newblock \bibinfo{title}{Cover cracking as a function of bar corrosion: Part
  {I}-{E}xperimental test}.
\newblock \bibinfo{journal}{Materials and Structures} \bibinfo{volume}{26},
  \bibinfo{pages}{453--464}.
\bibitem[{Atkinson and Thiercelin(1993)}]{AtkThi93}
\bibinfo{author}{Atkinson, C.}, \bibinfo{author}{Thiercelin, M.},
  \bibinfo{year}{1993}.
\newblock \bibinfo{title}{The interaction between the wellbore and
  pressure-induced fractures}.
\newblock \bibinfo{journal}{International Journal of Fracture}
  \bibinfo{volume}{59}, \bibinfo{pages}{23--40}.
\bibitem[{Barry and Aldis(1992)}]{BarAld92}
\bibinfo{author}{Barry, S.I.}, \bibinfo{author}{Aldis, G.K.},
  \bibinfo{year}{1992}.
\newblock \bibinfo{title}{Flow-induced deformation from pressurized cavities in
  absorbing porous tissues}.
\newblock \bibinfo{journal}{Bulletin of Mathematical Biology}
  \bibinfo{volume}{54}, \bibinfo{pages}{977--997}.
\bibitem[{Bellis et~al.(2017)Bellis, Vecchia, Ortiz and Pandolfi}]{deBVecOrt17}
\bibinfo{author}{Bellis, M.L.D.}, \bibinfo{author}{Vecchia, G.D.},
  \bibinfo{author}{Ortiz, M.}, \bibinfo{author}{Pandolfi, A.},
  \bibinfo{year}{2017}.
\newblock \bibinfo{title}{A multiscale model of distributed fracture and
  permeability in solids in all-round compression}.
\newblock \bibinfo{journal}{Journal of the Mechanics and Physics of Solids}
  \bibinfo{volume}{104}, \bibinfo{pages}{12--31}.
\bibitem[{de~Borst(1986)}]{Borst86}
\bibinfo{author}{de~Borst, R.}, \bibinfo{year}{1986}.
\newblock \bibinfo{title}{Non-linear analysis of frictional materials}.
\newblock Ph.D. thesis. Delft University of Technology.
  \bibinfo{address}{Delft, The Netherlands}.
\bibitem[{Cao et~al.(2018)Cao, Hussain and Schrefler}]{CaoHusSch18}
\bibinfo{author}{Cao, T.D.}, \bibinfo{author}{Hussain, F.},
  \bibinfo{author}{Schrefler, B.A.}, \bibinfo{year}{2018}.
\newblock \bibinfo{title}{Porous media fracturing dynamics: stepwise crack
  advancement and fluid pressure oscillations}.
\newblock \bibinfo{journal}{Journal of the Mechanics and Physics of Solids}
  \bibinfo{volume}{111}, \bibinfo{pages}{113--133}.
\bibitem[{Carrier and Granet(2012)}]{CarGra12}
\bibinfo{author}{Carrier, B.}, \bibinfo{author}{Granet, S.},
  \bibinfo{year}{2012}.
\newblock \bibinfo{title}{Numerical modeling of hydraulic fracture problem in
  permeable medium using cohesive zone model}.
\newblock \bibinfo{journal}{Engineering Fracture Mechanics}
  \bibinfo{volume}{79}, \bibinfo{pages}{312--328}.
\bibitem[{Chen et~al.(2000)Chen, Narayan, Yang and Rahman}]{CheNarYan00}
\bibinfo{author}{Chen, Z.}, \bibinfo{author}{Narayan, S.P.},
  \bibinfo{author}{Yang, Z.}, \bibinfo{author}{Rahman, S.S.},
  \bibinfo{year}{2000}.
\newblock \bibinfo{title}{An experimental investigation of hydraulic behaviour
  of fractures and joints in granitic rock}.
\newblock \bibinfo{journal}{International Journal of Rock Mechanics and Mining
  Sciences} \bibinfo{volume}{37}, \bibinfo{pages}{1061--1071}.
\bibitem[{Cheng(2016)}]{Che16}
\bibinfo{author}{Cheng, A.H.D.}, \bibinfo{year}{2016}.
\newblock \bibinfo{title}{Poroelasticity}. volume~\bibinfo{volume}{27}.
\newblock \bibinfo{publisher}{Springer}.
\bibitem[{Coussy(2010)}]{Cou11}
\bibinfo{author}{Coussy, O.}, \bibinfo{year}{2010}.
\newblock \bibinfo{title}{Mechanics and physics of porous solids}.
\newblock \bibinfo{publisher}{John Wiley \& Sons}.
\bibitem[{Cuss et~al.(2003)Cuss, Rutter and Holloway}]{CusRutHol03}
\bibinfo{author}{Cuss, R.J.}, \bibinfo{author}{Rutter, E.H.},
  \bibinfo{author}{Holloway, R.F.}, \bibinfo{year}{2003}.
\newblock \bibinfo{title}{Experimental observations of the mechanics of
  borehole failure in porous sandstone}.
\newblock \bibinfo{journal}{International Journal of Rock Mechanics and Mining
  Sciences} \bibinfo{volume}{40}, \bibinfo{pages}{747--761}.
\bibitem[{Damjanac et~al.(2016)Damjanac, Detournay and Cundall}]{DamDetCun15}
\bibinfo{author}{Damjanac, B.}, \bibinfo{author}{Detournay, C.},
  \bibinfo{author}{Cundall, P.A.}, \bibinfo{year}{2016}.
\newblock \bibinfo{title}{Application of particle and lattice codes to
  simulation of hydraulic fracturing}.
\newblock \bibinfo{journal}{Computational Particle Mechanics}
  \bibinfo{volume}{3}, \bibinfo{pages}{249--261}.
\bibitem[{Detournay(2004)}]{Det04}
\bibinfo{author}{Detournay, E.}, \bibinfo{year}{2004}.
\newblock \bibinfo{title}{Propagation regimes of fluid--driven fractures in
  impermeable rocks}.
\newblock \bibinfo{journal}{International Journal of Geomechanics}
  \bibinfo{volume}{4}, \bibinfo{pages}{35--45}.
\bibitem[{Detournay(2016)}]{Det16}
\bibinfo{author}{Detournay, E.}, \bibinfo{year}{2016}.
\newblock \bibinfo{title}{Mechanics of hydraulic fractures}.
\newblock \bibinfo{journal}{Annual Review of Fluid Mechanics}
  \bibinfo{volume}{48}, \bibinfo{pages}{311--339}.
\bibitem[{Detournay and Cheng(1995)}]{DetChe95}
\bibinfo{author}{Detournay, E.}, \bibinfo{author}{Cheng, A.H.D.},
  \bibinfo{year}{1995}.
\newblock \bibinfo{title}{Fundamentals of poroelasticity}, in:
  \bibinfo{booktitle}{Analysis and design methods}.
  \bibinfo{publisher}{Elsevier}, pp. \bibinfo{pages}{113--171}.
\bibitem[{Dresen et~al.(2010)Dresen, Stanchits and Rybacki}]{DreStaRyb10}
\bibinfo{author}{Dresen, G.}, \bibinfo{author}{Stanchits, S.},
  \bibinfo{author}{Rybacki, E.}, \bibinfo{year}{2010}.
\newblock \bibinfo{title}{Borehole breakout evolution through acoustic emission
  location analysis}.
\newblock \bibinfo{journal}{International Journal of Rock Mechanics and Mining
  Sciences} \bibinfo{volume}{47}, \bibinfo{pages}{426--435}.
\bibitem[{Fahy et~al.(2017)Fahy, Wheeler, Gallipoli and Grassl}]{FahWheGal17}
\bibinfo{author}{Fahy, C.}, \bibinfo{author}{Wheeler, S.},
  \bibinfo{author}{Gallipoli, D.}, \bibinfo{author}{Grassl, P.},
  \bibinfo{year}{2017}.
\newblock \bibinfo{title}{Corrosion induced cracking modelled by a coupled
  transport-structural approach}.
\newblock \bibinfo{journal}{Cement and Concrete Research} \bibinfo{volume}{94},
  \bibinfo{pages}{24--35}.
\bibitem[{Gale et~al.(2014)Gale, Laubach, Olson, Eichhubl and
  Fall}]{GalLauOls14}
\bibinfo{author}{Gale, J.}, \bibinfo{author}{Laubach, S.E.},
  \bibinfo{author}{Olson, J.E.}, \bibinfo{author}{Eichhubl, P.},
  \bibinfo{author}{Fall, A.}, \bibinfo{year}{2014}.
\newblock \bibinfo{title}{Natural fractures in shale: A review and new
  observationsnatural fractures in shale: A review and new observations}.
\newblock \bibinfo{journal}{AAPG bulletin} \bibinfo{volume}{98},
  \bibinfo{pages}{2165--2216}.
\bibitem[{Gale et~al.(2007)Gale, Reed and Holder}]{GalReeHol07}
\bibinfo{author}{Gale, J.F.W.}, \bibinfo{author}{Reed, R.M.},
  \bibinfo{author}{Holder, J.}, \bibinfo{year}{2007}.
\newblock \bibinfo{title}{Natural fractures in the barnett shale and their
  importance for hydraulic fracture treatments}.
\newblock \bibinfo{journal}{AAPG bulletin} \bibinfo{volume}{91},
  \bibinfo{pages}{603--622}.
\bibitem[{Goulty(2005)}]{Gou05}
\bibinfo{author}{Goulty, N.R.}, \bibinfo{year}{2005}.
\newblock \bibinfo{title}{Emplacement mechanism of the {Great Whin} and
  {Midland Valley} dolerite sills}.
\newblock \bibinfo{journal}{Journal of the Geological Society}
  \bibinfo{volume}{162}, \bibinfo{pages}{1047--1056}.
\bibitem[{Grassl and Bolander(2016)}]{GraBol16}
\bibinfo{author}{Grassl, P.}, \bibinfo{author}{Bolander, J.},
  \bibinfo{year}{2016}.
\newblock \bibinfo{title}{Three-dimensional network model for coupling of
  fracture and mass transport in quasi-brittle geomaterials}.
\newblock \bibinfo{journal}{Materials} \bibinfo{volume}{9},
  \bibinfo{pages}{782}.
\bibitem[{Grassl et~al.(2015)Grassl, Fahy, Gallipoli and Wheeler}]{GraFahGal15}
\bibinfo{author}{Grassl, P.}, \bibinfo{author}{Fahy, C.},
  \bibinfo{author}{Gallipoli, D.}, \bibinfo{author}{Wheeler, S.J.},
  \bibinfo{year}{2015}.
\newblock \bibinfo{title}{On a {2D} hydro-mechanical lattice approach for
  modelling hydraulic fracture}.
\newblock \bibinfo{journal}{Journal of the Mechanics and Physics of Solids}
  \bibinfo{volume}{75}, \bibinfo{pages}{104--118}.
\bibitem[{Jir\'{a}sek and Zimmermann(1998)}]{JirZim98}
\bibinfo{author}{Jir\'{a}sek, M.}, \bibinfo{author}{Zimmermann, T.},
  \bibinfo{year}{1998}.
\newblock \bibinfo{title}{Rotating crack model with transition to scalar
  damage}.
\newblock \bibinfo{journal}{Journal of Engineering Mechanics}
  \bibinfo{volume}{124}, \bibinfo{pages}{277--284}.
\bibitem[{Klinsmann et~al.(2016)Klinsmann, Rosato, Kamlah and
  McMeeking}]{KliRosKam16}
\bibinfo{author}{Klinsmann, M.}, \bibinfo{author}{Rosato, D.},
  \bibinfo{author}{Kamlah, M.}, \bibinfo{author}{McMeeking, R.M.},
  \bibinfo{year}{2016}.
\newblock \bibinfo{title}{Modeling crack growth during {Li} insertion in
  storage particles using a fracture phase field approach}.
\newblock \bibinfo{journal}{Journal of the Mechanics and Physics of Solids}
  \bibinfo{volume}{92}, \bibinfo{pages}{313--344}.
\bibitem[{Ladanyi(1967)}]{Lad67}
\bibinfo{author}{Ladanyi, B.}, \bibinfo{year}{1967}.
\newblock \bibinfo{title}{Expansion of cavities in brittle media}, in:
  \bibinfo{booktitle}{International Journal of Rock Mechanics and Mining
  Sciences \& Geomechanics Abstracts}, pp. \bibinfo{pages}{301--328}.
\bibitem[{Lecampion(2012)}]{Lec12}
\bibinfo{author}{Lecampion, B.}, \bibinfo{year}{2012}.
\newblock \bibinfo{title}{Modeling size effects associated with tensile
  fracture initiation from a wellbore}.
\newblock \bibinfo{journal}{International Journal of Rock Mechanics and Mining
  Sciences} \bibinfo{volume}{56}, \bibinfo{pages}{67--76}.
\bibitem[{Lecampion and Desroches(2015)}]{LecDes15}
\bibinfo{author}{Lecampion, B.}, \bibinfo{author}{Desroches, J.},
  \bibinfo{year}{2015}.
\newblock \bibinfo{title}{Simultaneous initiation and growth of multiple radial
  hydraulic fractures from a horizontal wellbore}.
\newblock \bibinfo{journal}{Journal of the Mechanics and Physics of Solids}
  \bibinfo{volume}{82}, \bibinfo{pages}{235--258}.
\bibitem[{McTigue(1987)}]{McT87}
\bibinfo{author}{McTigue, D.F.}, \bibinfo{year}{1987}.
\newblock \bibinfo{title}{Elastic stress and deformation near a finite
  spherical magma body: resolution of the point source paradox}.
\newblock \bibinfo{journal}{Journal of Geophysical Research: Solid Earth}
  \bibinfo{volume}{92}, \bibinfo{pages}{12931--12940}.
\bibitem[{van~der Meer et~al.(2009)van~der Meer, Kj{\ae}r, Kr{\"u}ger,
  J.~Rabassa and Kilfeather}]{Mee09}
\bibinfo{author}{van~der Meer, J.J.M.}, \bibinfo{author}{Kj{\ae}r, K.H.},
  \bibinfo{author}{Kr{\"u}ger, J.}, \bibinfo{author}{J.~Rabassa, J.},
  \bibinfo{author}{Kilfeather, A.A.}, \bibinfo{year}{2009}.
\newblock \bibinfo{title}{Under pressure: clastic dykes in glacial settings}.
\newblock \bibinfo{journal}{Quaternary Science Reviews} \bibinfo{volume}{28},
  \bibinfo{pages}{708--720}.
\bibitem[{Miehe et~al.(2015)Miehe, Mauthe and Teichtmeister}]{MieMauTei15}
\bibinfo{author}{Miehe, C.}, \bibinfo{author}{Mauthe, S.},
  \bibinfo{author}{Teichtmeister, S.}, \bibinfo{year}{2015}.
\newblock \bibinfo{title}{Minimization principles for the coupled problem of
  {D}arcy--{B}iot--type fluid transport in porous media linked to phase field
  modeling of fracture}.
\newblock \bibinfo{journal}{Journal of the Mechanics and Physics of Solids}
  \bibinfo{volume}{82}, \bibinfo{pages}{186--217}.
\bibitem[{Pantazopoulou and Papoulia(2001)}]{PanPap01}
\bibinfo{author}{Pantazopoulou, S.J.}, \bibinfo{author}{Papoulia, K.D.},
  \bibinfo{year}{2001}.
\newblock \bibinfo{title}{Modeling cover-cracking due to reinforcement
  corrosion in {RC} structures}.
\newblock \bibinfo{journal}{Journal of Engineering Mechanics}
  \bibinfo{volume}{127}, \bibinfo{pages}{342--351}.
\bibitem[{Rots(1988)}]{Rots88}
\bibinfo{author}{Rots, J.G.}, \bibinfo{year}{1988}.
\newblock \bibinfo{title}{Computational modeling of concrete fracture}.
\newblock Ph.D. thesis. Delft University of Technology.
  \bibinfo{address}{Delft, The Netherlands}.
\bibitem[{Savitski and Detournay(2002)}]{SavDet02}
\bibinfo{author}{Savitski, A.A.}, \bibinfo{author}{Detournay, E.},
  \bibinfo{year}{2002}.
\newblock \bibinfo{title}{Propagation of a penny--shaped fluid-driven fracture
  in an impermeable rock{: a}symptotic solutions}.
\newblock \bibinfo{journal}{International Journal of Solids and Structures}
  \bibinfo{volume}{39}, \bibinfo{pages}{6311--6337}.
\bibitem[{Selvadurai and Suvorov(2014)}]{SelSuv14}
\bibinfo{author}{Selvadurai, A.P.S.}, \bibinfo{author}{Suvorov, A.P.},
  \bibinfo{year}{2014}.
\newblock \bibinfo{title}{Thermo-poromechanics of a fluid-filled cavity in a
  fluid-saturated geomaterial}.
\newblock \bibinfo{journal}{Proceedings of the Royal Society A: Mathematical,
  Physical and Engineering Sciences} \bibinfo{volume}{470},
  \bibinfo{pages}{20130634}.
\bibitem[{Slowik and Saouma(2000)}]{SloSao00}
\bibinfo{author}{Slowik, V.}, \bibinfo{author}{Saouma, E.V.},
  \bibinfo{year}{2000}.
\newblock \bibinfo{title}{Water pressure in propagating concrete cracks}.
\newblock \bibinfo{journal}{Journal of Structural Engineering}
  \bibinfo{volume}{126}, \bibinfo{pages}{235--242}.
\bibitem[{Stanchits et~al.(2011)Stanchits, Mayr, Shapiro and
  Dresen}]{StaMaySha11}
\bibinfo{author}{Stanchits, S.}, \bibinfo{author}{Mayr, S.},
  \bibinfo{author}{Shapiro, S.}, \bibinfo{author}{Dresen, G.},
  \bibinfo{year}{2011}.
\newblock \bibinfo{title}{Fracturing of porous rock induced by fluid
  injection}.
\newblock \bibinfo{journal}{Tectonophysics} \bibinfo{volume}{503},
  \bibinfo{pages}{129--145}.
\bibitem[{Tarokh et~al.(2016)Tarokh, Blanksma, Fakhimi and Labuz}]{TarBlaFak16}
\bibinfo{author}{Tarokh, A.}, \bibinfo{author}{Blanksma, D.J.},
  \bibinfo{author}{Fakhimi, A.}, \bibinfo{author}{Labuz, J.F.},
  \bibinfo{year}{2016}.
\newblock \bibinfo{title}{Fracture initiation in cavity expansion of rock}.
\newblock \bibinfo{journal}{International Journal of Rock Mechanics and Mining
  Sciences} \bibinfo{volume}{85}, \bibinfo{pages}{84--91}.
\bibitem[{Timoshenko and Goodier(1987)}]{TimGoo87}
\bibinfo{author}{Timoshenko, S.P.}, \bibinfo{author}{Goodier, J.N.},
  \bibinfo{year}{1987}.
\newblock \bibinfo{title}{Theory of elasticity}.
\newblock \bibinfo{publisher}{McGraw-Hill}, \bibinfo{address}{New York}.
\bibitem[{Viesca and Garagash(2018)}]{VieGar18}
\bibinfo{author}{Viesca, R.C.}, \bibinfo{author}{Garagash, D.I.},
  \bibinfo{year}{2018}.
\newblock \bibinfo{title}{Numerical methods for coupled fracture problems}.
\newblock \bibinfo{journal}{Journal of the Mechanics and Physics of Solids}
  \bibinfo{volume}{113}, \bibinfo{pages}{13--34}.
\bibitem[{Vlahou and Worster(2010)}]{VlaWor10}
\bibinfo{author}{Vlahou, I.}, \bibinfo{author}{Worster, M.G.},
  \bibinfo{year}{2010}.
\newblock \bibinfo{title}{Ice growth in a spherical cavity of a porous medium}.
\newblock \bibinfo{journal}{Journal of Glaciology} \bibinfo{volume}{56},
  \bibinfo{pages}{271--277}.
\bibitem[{Wilson and Landis(2016)}]{WilLan16}
\bibinfo{author}{Wilson, Z.A.}, \bibinfo{author}{Landis, C.M.},
  \bibinfo{year}{2016}.
\newblock \bibinfo{title}{Phase-field modeling of hydraulic fracture}.
\newblock \bibinfo{journal}{Journal of the Mechanics and Physics of Solids}
  \bibinfo{volume}{96}, \bibinfo{pages}{264--290}.
\bibitem[{Xing et~al.(2017)Xing, Yoshioka, Adachi, El-Fayoumi and
  Bunger}]{XinYosAda17}
\bibinfo{author}{Xing, P.}, \bibinfo{author}{Yoshioka, K.},
  \bibinfo{author}{Adachi, J.}, \bibinfo{author}{El-Fayoumi, A.},
  \bibinfo{author}{Bunger, A.P.}, \bibinfo{year}{2017}.
\newblock \bibinfo{title}{Laboratory measurement of tip and global behavior for
  zero-toughness hydraulic fractures with circular and blade-shaped {(PKN)}
  geometry}.
\newblock \bibinfo{journal}{Journal of the Mechanics and Physics of Solids}
  \bibinfo{volume}{104}, \bibinfo{pages}{172--186}.
\bibitem[{Yu and Houlsby(1991)}]{YuHou91}
\bibinfo{author}{Yu, H.S.}, \bibinfo{author}{Houlsby, G.T.},
  \bibinfo{year}{1991}.
\newblock \bibinfo{title}{Finite cavity expansion in dilatant soils: loading
  analysis}.
\newblock \bibinfo{journal}{Geotechnique} \bibinfo{volume}{41},
  \bibinfo{pages}{173--183}.
\bibitem[{Zhang et~al.(2011)Zhang, Jeffrey, Bunger and
  Thiercelin}]{ZhaJefBun11}
\bibinfo{author}{Zhang, X.}, \bibinfo{author}{Jeffrey, R.G.},
  \bibinfo{author}{Bunger, A.P.}, \bibinfo{author}{Thiercelin, M.},
  \bibinfo{year}{2011}.
\newblock \bibinfo{title}{Initiation and growth of a hydraulic fracture from a
  circular wellbore}.
\newblock \bibinfo{journal}{International Journal of Rock Mechanics and Mining
  Sciences} \bibinfo{volume}{48}, \bibinfo{pages}{984--995}.

\end{thebibliography}

\end{document}